\documentclass[a4paper,11pt]{article}
\pdfoutput=1
\usepackage{jheppub}
\usepackage{subcaption}
\allowdisplaybreaks
\usepackage{xspace}
\usepackage{enumitem}

\renewcommand{\Re}{\operatorname{Re}}

\newcommand{\MSb}{\ensuremath{\overline{\text{MS}}}\xspace}
\newcommand{\eeggs}{\ensuremath{e^+e^-\to\gamma\gamma^*}\xspace}
\newcommand{\aMS}{\bar{a}}
\newcommand{\e}{\epsilon}
\newcommand{\tLmu}{\widetilde{L}_\mu}
\newcommand{\tLm}{\widetilde{L}}
\newcommand{\Lmu}{L_\mu}
\newcommand{\Lm}{L}

\newcommand{\be}{\begin{equation}}
\newcommand{\ee}{\end{equation}}
\setlist[itemize,1]{label=$\triangleright$}
\setlist[itemize,2]{label=$\bullet$}
\newcommand{\alMS}{\alpha_{\MSb}}

\title{Two-loop radiative corrections to $e^+ e^-\rightarrow \gamma\gamma^*$ cross section.}

\author{V.S. Fadin}
\author{and R.N. Lee}

\affiliation{Budker Institute of Nuclear Physics, Novosibirsk 630090, Russia}

\emailAdd{v.s.fadin@inp.nsk.su}
\emailAdd{r.n.lee@inp.nsk.su}

\abstract{The increasing accuracy of current and planned experiments to measure the anomalous magnetic moment of the muon requires  more precision and reliability of  its theoretical calculation. For this purpose, we calculate the differential cross section for the process of annihilation of an electron-positron pair into two photons, one of which is virtual, accompanied by the emission of soft photons, taking into account radiative corrections of the order $\alpha^2$.   The results obtained can be used to improve the accuracy of calculating the contribution of the hadron vacuum polarization to the muon anomalous moment. It is shown that all logarithmically amplified two-loop corrections can be easily found using modern theorems of soft and collinear factorizations and available one-loop results.}

\begin{document}

\maketitle
\flushbottom

\section{Introduction}
At the present time, the entire set of experimental data obtained on man-made physical devices is fairly well described by the so-called Standard Model (SM) of strong and electroweak interactions (with the inclusion of right-handed neutrinos). It is clear, however, that the model has a limited range of applicability. It does not meet both the requirements of a high theory and the questions of astrophysics and cosmology. Therefore, the main goal of modern high-energy physics is proclaimed to be the detection of processes and phenomena that are not described by the Standard Model, i.e. this very New Physics.

The discrepancy of 4.2 standard deviations between the experimentally measured \cite{Muong-2:2006rrc, Muong-2:2021ojo} muon anomalous magnetic moment (MAMM) and its theoretically obtained SM  value \cite{Aoyama:2020ynm} is one of   the most significant deviations of the experiment from the SM predictions\footnote{Very recently the new experimental result for MAMM was published in Ref. \cite{Aguillard2023}. This result agrees with those of Refs.  \cite{Muong-2:2006rrc, Muong-2:2021ojo}, being significantly much more accurate.}.
%, yielding in size only to the recently published \cite{CDF:2022hxs} discrepancy of 7 standard deviations between the experimentally measured W-boson mass and its theoretical value
A possible (and currently most discussed) explanation of this discrepancy is the contribution to the MAMM of New Physics, that is, particles and interactions not represented in the SM. Often this discrepancy is even considered as undeniable evidence of New Physics. Under these conditions, the verification of SM predictions and an increase in the accuracy of these predictions are of particular importance.

The inaccuracy of predictions is mainly due to the fact that the SM contains contributions to the MAMM, which cannot be obtained ``from first principles.'' These are hadronic contributions. They  are divided into two types: those coming from the vacuum hadronic polarization and from the hadronic contribution to the scattering of light by light (to the amplitude of off-shell photon-photon scattering). The largest of them is the contribution from the vacuum hadronic polarization (it is almost two orders of magnitude larger than the contribution from photon-photon scattering). The predictions of this contribution are not entirely theoretical, since they are obtained using experimental data on the annihilation of an electron-positron pair into hadrons. But there is no ``pure'' annihilation into hadrons. Experimentally measured cross sections contain radiative corrections associated with electromagnetic interaction. Therefore, the accuracy of  ``theoretical'' calculations  of the contribution from  the vacuum hadronic polarization depends not only on the accuracy of experiment, but also on the accuracy of calculations of the radiative corrections. Moreover, it is this contribution that makes the greatest uncertainty in the theoretical value of the MAMM.

The   contribution of the vacuum polarization by hadrons is expressed in terms of the quantity
\begin{equation}\label{eq:Rratio}
	R(s) =\sigma( e^+e^-\rightarrow \gamma^*\rightarrow {\textit{hadrons}})/\sigma^{(0)}({e^+e^-\rightarrow \mu^+\mu^-}),
\end{equation}
i.e. the ratio of the inclusive cross section of the one-photon annihilation of an electron-positron pair with total energy $\sqrt s $ into hadrons (that is, the cross section of the process in which at least one hadron is observed in the final state) to the total annihilation cross section into the $\mu^+\mu^-$ pair, calculated in the Born approximation in neglecting the masses of all particles. It is this quantity that enters in the dispersion representation for the contribution of hadronic vacuum polarization to the MAMM \cite{Bouchiat:1961lbg, Brodsky:1967sr}. The inclusive cross section can be measured only in the region of relatively large energies $\sqrt s\gtrsim 1.8$ GeV. Meanwhile, it is the region of lower energies that dominates the dispersion integral for the MAMM. In this region the inclusive cross section is obtained as a sum of measured exclusive cross sections for different channels, which provides a much better accuracy than the direct measurement of this quantity. Nevertheless, it turns out that the uncertainty of this contribution in the  prediction of the MAMM is the largest (see, for example, the review \cite{Logashenko:2018pcl}). Therefore, the requirements for the accuracy of its extraction from experimental data should be the most stringent.

A correct assessment of the accuracy of the theoretical prediction for the hadronic vacuum polarization contribution to the MAMM  and its improvement has become even more important after its recent lattice calculations \cite{Borsanyi:2020mff,Toth:2022lsa} with declared accuracy of 0.8\%. The result of these calculations turns out to be much closer to experiment and reduces the discrepancy between experiment and theory to only $1.5$ standard deviations.

And finally, the importance of the correct account for the radiative corrections increased even more after the appearance of the paper \cite{CMD-3:2023alj}, in which the measured cross section of the process $e^+e^-\rightarrow  \gamma^* \rightarrow\pi^+\pi^-$ was used to estimate the contribution $\pi^+\pi^-$ to hadronic part of the MAMM in the energy range
$0.6\text{ GeV} < \sqrt s < 0.88 \text{ GeV}$. The value based on the CMD-3 data is significantly larger than the estimates based on the results of previous measurements, and substantially reduces the discrepancy between the experimental value of the MAMM and its prediction in the Standard Model.

The accuracy of extracting the value of $R(s)$ from the experimental data on electron-positron annihilation into hadrons depends critically on the accuracy of taking into account the radiative corrections to the annihilation cross section. Despite the smallness of the coupling constant $\alpha \approx 1/137.036$ in quantum electrodynamics (QED), in order to achieve a high (fraction of a percent) accuracy, the summation of the contributions of several terms of the perturbation theory is required, since powers of $\alpha$  are followed by powers of logarithms
$\ln(s/m^2)$ and $\ln(E/\Delta E)$, where  $m$  is the electron mass,  $E =\sqrt s/2$ and $\Delta  E$ is the  maximum energy loss for radiation allowed by the experimental conditions.

Currently, there are  two types of experiments  to measure the cross sections for annihilation of an electron-positron pair into hadrons: energy scan measurements and radiation return measurements.   The second type experiments  became possible with the appearance of high-energy colliders with high luminosity, which makes it possible to collect statistics comparable to those achievable in experiments of the first type when measuring cross sections suppressed in the fine structure constant.  This method   is widely used now for obtaining of the cross section $\sigma_{e^+e^-\rightarrow hadrons}$ in the entire region important for calculation of the hadronic contribution to $a_\mu =(g_\mu -2)/2$.

An efficient method for summing radiative corrections enhanced by $\ln(s/m^2)$ is the method of parton distributions (also called the method of structure functions), developed for computing radiative corrections to the single-photon $e^+e^-$ annihilation cross section in Ref. \cite{kuraev1985radiative} by analogy with the calculation of the Drell-Yan process cross section in QCD.
\footnote{The analogy with QCD was used earlier for calculation of QED radiative corrections to quasielastic neutrino scattering in Ref. \cite{de1979radiative}.}.
However,  this method  provides the best accuracy when calculating radiative corrections to the inclusive cross section for hadron production in energy scan experiments. When calculating the radiative corrections to the cross sections for exclusive hadron production, the accuracy of this method decreases due to the fact that parton distributions are calculated without taking into account the restrictions imposed on the kinematics of emitted particles by the event selection criteria. The accuracy is also reduced when applying this method to experiments with radiative return due to the presence of additional parameters associated with the emission of a photon that provides this return. All these circumstances make it extremely desirable to directly calculate the radiative corrections.

In this paper we calculate in the  two-loop approximation  the cross section of the process of  electron-positron annihilation into a pair of photons, one of which is virtual. Although the term cross section  is not quite correct  when applied to a virtual particle, we use it because the quantity being calculated has all the properties of a cross section.
The  cross section of the process $e^+e^-\rightarrow  \gamma +\gamma^*\rightarrow  \gamma +\textit{hadrons}$ is obtained from this quantity by convolution with hadronic tensor, as described below.
The results obtained can be used  when processing data in experiments of both types: by the radiation return method and by the energy scanning method.
In the first case it gives directly  the cross section of one-photon hadron production, taking into account two-loop radiative corrections associated with the interaction of the initial electron-positron pair, differential in angle and energy of the photon, which ensures the radiative return.
In the second case, in order to obtain the corresponding radiative corrections, the integration over the angles and energies of this photon should be carried out.

It should be noted here that, in addition to the above mentioned radiative corrections, the experimentally measured cross sections also include corrections due to the interaction in the final state and those related to the hadron production mechanisms different from the one-photon one. In general, these corrections depend on the structure of hadrons, so their \textit{ab initio} calculation is hardly possible. Their  account  requires special consideration and goes far beyond the scope of the work under consideration.
Here we confine ourselves to the remark that the experimental setup symmetric in the sign of the hadron charge (as in the measurement of inclusive cross sections) is preferable, since it is less sensitive to the amplitudes of two-photon production of hadrons, which depend on their structure. As was shown in Ref. \cite{Ignatov:2022iou}, taking this structure into account is absolutely important for describing the experimental data on the charge asymmetry in the  process $e^+e^-\rightarrow \pi^+\pi^-$. However, in a charge-symmetric setup of the experiment, the interference of one-photon and two-photon production mechanisms does not contribute, so that the corrections associated with other than one-photon hadron production mechanisms appear only in two loops (in the order of alpha squared). Since the corresponding amplitudes do not have collinear singularities, and infrared singularities are canceled in the inclusive cross section, the relative magnitude of the corresponding radiative corrections can be estimated as $\left(\tfrac{\alpha}{\pi}\right)^2$. It limits the accuracy of model-independent calculation of radiative corrections.
Note that for our present results this accuracy restriction means that only the terms amplified by large logarithms $\ln(E/m)$ and $\ln(E/\Delta E)$ are model-independent.

%Note that our results for the differential cross section have the form
%\begin{equation}
%	a_0+a_1 L+a_2 L^2+a_3 L^3+a_4 L_{\omega }+a_5 L L_{\omega }+a_6 L^2 L_{\omega }+a_7 L_{\omega }^2+a_8 L L_{\omega }^2+a_9 L^2 L_{\omega }^2
%\end{equation}

The paper is organized as follows. In Section 2 we discuss several factorization properties of the  QED amplitudes: factorization of soft and collinear singularities in massless and massive QED, relation between massless and massive QED amplitudes, and factorization of soft radiation in inclusive cross section. In Section 3 we underline some consequences of the factorization properties. In particular, we show that all logarithmically amplified terms in N$^l$LO contribution to the amplitude or cross section are expressed in terms of N$^k$LO contributions with $k<l$.
Some details of calculation are given in Section 4 and the results are presented in Section 5.  Section 6 contains the Conclusion.

%TODO: consistently use either infrared or soft.

\section{Factorization of QED  amplitudes and cross sections}

The sources of the logarithms  $\ln(E/m)$ and $\ln(E/\Delta E)$ are infrared and collinear  singularities in perturbation theory. It is known that these singularities factorize. The factorization of the infrared singularities in QED is well known since \cite{Bloch:1937pw,Yennie:1961ad}. The theory of factorization of collinear singularities, while being in its infancy in early works on QED \cite{Kessler:1962ffa,Baier:1973ms}, later was well developed in quantum chromodynamics (QCD) \cite{Catani:1998bh, Sterman:2002qn, Dixon:2008gr,  Aybat:2006wq, Becher:2007cu, Becher2009, Becher:2009cu, Gardi2009, Gardi2009a}.

\subsection{Soft-collinear factorization in massless QED}

The dimensional regularization and minimum subtraction (\MSb) renormalization scheme is the most widely used approach for perturbative calculations in QCD. Within this approach the infrared and collinear singularities appear as poles in $\e = (4-D)/2$, where $D$  is the space-time dimension, taken to be different from the physical value $D=4$.
The soft-collinear factorization means that the massless amplitude $\mathcal{A}$ can be written as a product of $\e$-singular factor  $\mathcal{Z}$, containing information about infrared and collinear singularities, and the ``hard'' amplitude $\mathcal{H}$, finite  at  $\e = 0$. In QCD both $\mathcal{Z}$ and  $\mathcal{H}$ are matrices in the color space. In QED, thanks to the absence of color indices, the situation is greatly simplified and we have
\begin{equation}
%\mathcal{A}(\{p\}, \e, \mu,  \aMS) =\mathcal{Z}(\{p\}, \e, \mu,  \aMS) \mathcal{H}(\{p\}, \e, \mu,  \aMS)\,, \label{eq:massless_factorization}
\mathcal{A}(in\to out) =\mathcal{Z}(in\to out) \mathcal{H}(in\to out)\,, \label{eq:massless_factorization}
\end{equation}
where (cf. Eq. (2.9) of Ref. \cite{Becher2009})
\begin{multline}
%	\mathcal{Z}(\{p\}, \e, \mu, \aMS)
	\mathcal{Z}(in\to out)
	=\exp\bigg\{\int_0^{\aMS} \tfrac{d a_1}{a_1\beta(\e, a_1)} \bigg[-\tfrac12 \sum_{i}\gamma_i(a_1)\\
	+\tfrac14 \sum_{i<j} Q_iQ_j
	\bigg(\gamma_{K}(a_1)\ln \left(\tfrac{-(p_i+p_j)^2-i0}{\mu^2}\right)  +\int_0^{a_1} \tfrac{d a_2\,\gamma_{K}(a_2)}{a _2\beta(\e, a_2)} \bigg)
	\bigg] \bigg\}\,. \label{eq:Z_massless}
\end{multline}
The double sum $\sum_{i<j}$ in Eq. \eqref{eq:Z_massless} runs over all pairs of particles, thus this formula is called ``dipole'' in some literature. Note that the corresponding formula in QCD, apart from acquiring obvious modifications --- retaining the color indices in $\gamma_i$ and $Q_i$ and replacing $\exp$ with path-ordered exponent --- must be supplemented by contributions of higher ``multipoles''. These are contributions containing sums over $k$-tuples of particles with $k>2$. The  quadrupole correction first appearing in three loops was calculated relatively recently \cite{Almelid:2015jia}.

In \eqref{eq:massless_factorization} and \eqref{eq:Z_massless} we used the following notations: $\{p\}=\{p_1,\ldots,p_n\}$ denotes the set of momenta of external particles (all momenta are considered as incoming), $\mu$ is the renormalization scale,  $\aMS = \aMS (\mu) = \alMS(\mu)/(4\pi)$, $\alMS(\mu)=e_{\MSb}^2(\mu)/(4\pi)$ is the \MSb renormalized QED coupling constant,  $\beta(\e, a)=\tfrac{d\ln \aMS}{d\ln \mu^2}$ is the beta function in $D= 4-2\e$ dimensions, $\gamma_{K}(\aMS)$  is the light-like cusp anomalous dimension (defined as a slope of cusp anomalous dimension at large angle), $p_i$ and $\gamma_i(\aMS)$ are the momentum and collinear dimension of $i$-th particle, respectively. The sign variable $Q_i$ equals to $1$ for incoming electron or outgoing positron, to $-1$ for incoming positron or outgoing electron, and to $0$ for photon.

The perturbative expansions of $\beta$, $\gamma_K$, and $\gamma_i$ entering Eq. \eqref{eq:Z_massless} have the form

\begin{equation}\label{eq:pertexp}
	\beta(\e, \aMS) = - \e -\sum_{l=0}^\infty\beta_l \aMS^{l+1},
	\quad \gamma_K(\aMS) = \sum_{l=1}^{\infty} \gamma_K^{(l)}(4\aMS)^l,
	\quad \gamma_i(\aMS) = \sum_{l=1}^{\infty} \gamma_i^{(l)}\bar{a}^l\,.
\end{equation}

For our purposes we need to know  the expansion of Eq. \eqref{eq:Z_massless} only up to $\aMS^2$:

\begin{multline}
%	\ln\mathcal{Z}(\{p\}, \e, \mu, \aMS)
	\ln\mathcal{Z}(in \to out)
	= \aMS \Bigg[ \sum_{i<j} Q_iQ_j \tfrac{\gamma^{(1)}_{K}}{\e} \Bigg(-\ln \left(\tfrac{-(p_i+p_j)^2-i0}{\mu^2}\right)+\tfrac{1}{\e}\Bigg) +\tfrac12\sum_{i}\tfrac{\gamma^{(1)}_i}{\e} \Bigg]
\\
+\aMS^2
\Bigg[ \sum_{i<j} Q_iQ_j  \Big(\tfrac{\gamma^{(1)}_{K} \beta_0}{2\e^2} -2\tfrac{\gamma^{(2)}_{K}}{\e}\Big) \ln \left(\tfrac{-(p_i+p_j)^2-i0}{\mu^2}\right) - \tfrac{3\gamma^{(1)}_{K} \beta_0}{4\e^3} +\tfrac{\gamma^{(2)}_{K}}{\e^2}
\\
+\tfrac14\sum_{i}\left(-\tfrac{\gamma^{(1)}_i \beta_0}{\e^2} + \tfrac{\gamma^{(2)}_i}{\e} \right)\Bigg]
+\mathcal{O}(\aMS^3)\,.\label{eq:Z_massless_1}
\end{multline}
The required expansion coefficients  can be extracted from the corresponding QCD results available in the literature \cite{Korchemsky:1985xj, Korchemsky:1987wg, %Dixon:2008gr,
Becher:2009cu}. We have
\begin{align}
	\gamma^{(1)}_{K} &= 2\,,
	&\quad\gamma^{(2)}_{K} &= -\tfrac{10}{9}n_f\,,
\\
\gamma^{(1)}_{e} &= -3\,,
&\gamma^{(2)}_{e} &= -\tfrac32 +12\zeta_2 -24\zeta_3+\left(\tfrac{130}{27} +4\zeta_2\right)n_f\,,
\\
\gamma^{(1)}_{\gamma} &= -\beta_0=\tfrac43 n_f\,,
&\gamma^{(2)}_{\gamma} &= -\beta_1=4n_f~. \label{betagamma coefficients}
\end{align}
Here and below $n_f$ stands for the number of fermions, equal to 1 in our calculations, but we keep it as a symbolic parameter to trace  the contributions from electronic loops.

Now we want to specialize Eq. \eqref{eq:Z_massless_1} to the process of our present interest, \begin{equation}
	e^-(p_-) +e^+(p_+)\rightarrow\gamma(k) + \gamma^*(q)\,.
\end{equation}
We assume that the summation indices in \eqref{eq:Z_massless_1}  run over the set $\{e^+,e^-,\gamma\}$, not including the off-shell photon $\gamma^*$. We obtain
\begin{multline}
	\ln\mathcal{Z}(e^+ e^- \rightarrow \gamma\gamma^*)=
	%calZ
	\aMS\, \Big[-\tfrac{2}{\e ^2}+\tfrac{2 \tLmu-3}{\e}\underline{+\tfrac{2n_f}{3\e}}\Big]
	+\aMS^2\, \Big[-\tfrac{3 - 24 \zeta_2 + 48\zeta_3}{4\e }
	\\
	+\left(-\tfrac{2}{\e ^3}-\tfrac{4(2-3 \tLmu)}{9\e ^2} + \tfrac{65+54 \zeta_2-60 \tLmu}{27\e}\right)n_f
	\underline{
		+\tfrac{4n_f^2}{9\e^2}
		+\tfrac{n_f}{\e}
	}\Big]
	+\mathcal{O}\left(\aMS^3\right)\,, \label{eq:cal_Z}
\end{multline}
where  $\tLmu=\ln[(-s-i0)/\mu^2], \;\; s=(p_-+p_+)^2$. The underlined terms come from the contributions $\propto \gamma_\gamma$.

The factorization property means that the amplitude of the process at zero electron mass, $\mathcal{A}_\mu(e^+ e^-\rightarrow \gamma\gamma^*)$, can be represented as
\begin{equation}
\mathcal{A}_\mu(e^+ e^- \rightarrow \gamma\gamma^*) = \mathcal{Z}(e^+ e^- \rightarrow \gamma\gamma^*) \cdot  \mathcal{H}_\mu(e^+ e^- \rightarrow \gamma\gamma^*)\,, \label{eq:massless_factorization_ggstar}
\end{equation}
where $\mathcal{Z}$ is defined in Eq. \eqref{eq:cal_Z} and $\mathcal{H}$ is finite at $\e=0$. Although we don't have any method of calculating the hard amplitude apart from direct calculation of $\mathcal{A}_\mu(e^+ e^- \rightarrow \gamma\gamma^*)$ and using Eq. \eqref{eq:massless_factorization_ggstar}, we shall see soon, that this factorization allows us to express all terms amplified by $\ln (E/m)$ and/or by $\ln (E/\Delta E)$ in NNLO cross section via one-loop results, which have a relatively simple form and are available in the literature.

A few remarks should be made here. First, the massless amplitude $\mathcal{A}_\mu(e^+ e^- \rightarrow \gamma\gamma^*)$ is understood as a perturbative expansion defined via diagrams with bare electric charge $e_0$ in all vertices except the vertex, corresponding to the off-shell photon, where we put $\gamma_\mu$ without any charge at all. We also amputate the external leg, corresponding to $\gamma^*$. We stress that no additional factors related to the wave function renormalization are included in $A_\mu(e^+ e^- \rightarrow \gamma\gamma^*)$. The ``renormalization'' of the amplitude is understood as plainly expressing all bare couplings $\alpha_0$ via $\alMS(\mu)$ with $\alpha_0 =\left(\tfrac{ \mu^2 e^\gamma}{4\pi}\right)^{\e}\alpha_{\MSb}(\mu)/ Z^{\MSb}_3$, see Eq. \eqref{eq:amsviaa0} in Appendix.

Note also that  the factorization formula \eqref{eq:massless_factorization} was derived in QCD, and for on-shell scattering amplitudes,  whereas we want to use it in QED, and for the amplitude $A_\mu$ with one off-shell external momentum $q$. Both differences,  between asymptotically free QCD and QED with zero charge problem, and between  on-shell and off-shell  amplitudes, prompted us to directly test the validity of the factorization. We have checked, up to two loops, that pulling out the factor $\mathcal{Z}$, Eq. \eqref{eq:cal_Z}, from $A_\mu$, as prescribed by Eq. \eqref{eq:massless_factorization_ggstar}, we obtain finite at $\e=0$ hard amplitude $\mathcal{H}_\mu$. Note that in this check it was essential to use physical polarization $e$ of the real photon, i.e. the identity $e\cdot k=0$, otherwise we observed $\propto \tfrac{e\cdot k}{\e}$ terms in $\mathcal{H}_\mu$ both in one- and in two-loop results.

%are valid in the two-loop approximation  for  the amplitude  $A(e^+ e^- \rightarrow \gamma\gamma^*) $ without  taking into account the vacuum polarization contributions to the virtual photon line. It was done in the following way.  First, the  bare (unrenormalized) amputated (without self-energy  corrections to the external lines) amplitude  $A(e_+ e_-\rightarrow \gamma\gamma^*)$  was calculated.  Then, the unrenormalized QED coupling $\alpha_0$ was expressed in terms of the renormalized one  in the ${\overline{MS}}$ scheme  using the relation
%\begin{equation}
%\alpha_0 =\left(\tfrac{ \mu^2 e^\gamma}{4\pi}\right)^{\e}\alpha_{\MSb}(\mu) \overline{Z}^{-1}_3\,,
%\quad
%\overline{Z}_3  =1  + \tfrac{\beta_0}{\e}\aMS +\tfrac{\beta_1{\CancelColor-\beta_0^2}}{2\e} \aMS ^2\,.
%%\bar{\mu}^2=\tfrac{ \mu^2 e^\gamma}{4\pi}, \;\;
%%\aMS = \tfrac{\alpha_{\MSb}(\mu)}{4\pi}\,,  \label{Z 3}
%\label{eq:a0viaams}
%\end{equation}
%Here $\gamma = 0, 5772157....$ is the Euler constant.
%After that it was checked that the product $A(e_+ e_-\rightarrow \gamma\gamma^*) \mathcal{Z}^{-1}$, where $\mathcal{Z}$ is defined by \eqref{eq:Z_massless_1} is not singular
%at $\e\rightarrow 0$. Note that the last statement is valid for amplitudes with physical polarization  of the real photon only.

\subsection{Relation between massive and massless amplitudes}

Another factorization  formula  based on the soft-collinear effective theory \cite{Bauer:2000yr, Beneke:2002ph}  was suggested in Ref. \cite{Becher:2007cu}. This formula relates massive QED amplitude ${A}(in\to out)$ and the corresponding massless amplitude $\mathcal{A}(in\to out)$. The relation can be written in the following form
\begin{equation}
	{A}(in\to out)
	=\left[Z_3^{\text{OS}}\right]^{k/2}\left[Z_J\right]^{n/2}S(in\to out)\mathcal{A}(in\to out)+\mathcal{O}(m^2/Q^2)\,.\label{eq:massive_factorization}
\end{equation}
Here $k$ is the number of external photon legs, $n$ is the number of external electron/positron legs,  $Z_J=Z_J(m^2)$ is the so called \textit{jet function}, depending only on the particle mass, and $S(in\to out)$ is the so called \textit{soft function} depending in a prescribed way on the kinematics of the process. Note that in Ref. \cite{Becher:2007cu} the above factorization formula was used for the case $k=0$, however, the presence of additional factor $\left[Z_3^{\text{OS}}\right]^{k/2}$ for $k\neq0$ seems to be quite obvious.

The soft function appears in massive QED  due to the fact that  integrals for the contributions of soft photons cease to be scale-free when the vacuum polarization is taken into account. It has the form
\begin{equation}
	S(in\to out)=\exp\left\{-\sum_{i<j} Q_iQ_j \delta S\left(-(p_i+p_j)^2-i0,m^2\right)\right\}\,,
\end{equation}
where (cf. \cite[Eq. (3.9)]{Becher:2007cu})

\begin{equation}
	\delta S(Q^2, m^2) = \aMS^2 \left(\tfrac{\mu}{m}\right)^{4\e} n_f\,\left[-\tfrac{4}{3 \e ^2}+\tfrac{20}{9 \e }-\tfrac{112}{27} -\tfrac{4}{3}\zeta_2+O\left(\e ^1\right)\right]\ln(Q^2/m^2) + \mathcal{O}\left(\aMS^3\right)\,. \label{eq:deltaS}
\end{equation}

The jet function $Z_J$ was determined in Ref. \cite{Becher:2007cu} from the factorization relation written for the Dirac form factor
\begin{equation}
	F(Q^2) = Z_J(m^2)\left[1+\delta S(Q^2, m^2)\right]\mathcal{F}(Q^2)+\mathcal{O}(m^2/Q^2), \label{eq:F=ZSF}
\end{equation}
where $Q^2 = -q^2$, $q$ is the momentum transfer, $F(Q^2)$ and $\mathcal{F}(Q^2)$ are the Dirac form factors in massive and massless QED, respectively. Explicit calculation of $F(Q^2)$ and $\mathcal{F}(Q^2)$ up to two loops allows one to obtain $Z_J$ with the corresponding precision.
The result can be written as
\begin{multline}
\ln Z_J=
\aMS \left(\tfrac{\mu}{m}\right)^{2\e} \left(Z^{\MSb}_3\right)^{-1}\, \left[\tfrac{2}{\e ^2}+\tfrac{1}{\e }+4+\zeta_2+\left(8+\tfrac{\zeta_2}{2}-\tfrac{2 \zeta_3}{3}\right) \e
%+\left(16+2 \zeta_2-\tfrac{\zeta_3}{3}+\tfrac{9 \zeta_4}{8}\right) \e ^2+\mathcal{O}\left(\e ^3\right)
+\mathcal{O}\left(\e ^2\right)\right]
\\
+\aMS^2 \left(\tfrac{\mu}{m}\right)^{4\e}\,\Big[\tfrac{3 - 24\zeta_2 + 48 \zeta_3}{4\e }+\tfrac{177}{8}+22 \zeta_2-48 \zeta_2\ln {2}-6 \zeta_3-42 \zeta_4
\\
+\left(-\tfrac{2}{3 \e ^3}-\tfrac{4}{9 \e ^2}-\tfrac{90 \zeta_2+209}{27\e }-\tfrac{76 \zeta_2}{9}+\tfrac{4 \zeta_3}{9}+\tfrac{3379}{162}\right)n_f
+\mathcal{O}\left(\e ^1\right)\Big]+\mathcal{O}\left(\aMS^3\right)\,.\label{eq:lnZ_J}
\end{multline}
The convenience of using $\ln Z_J$ is that in this quantity (but not in $Z_J$ itself), we can safely omit terms vanishing at $\e\to 0$. Note that we have kept in the first pair of brackets the $\e^1$ term because the factor $\left(Z^{\MSb}_3\right)^{-1}=1-\tfrac{\beta_0}{\e}\aMS+O(\aMS^2)$ has simple pole, cf. \eqref{eq:amsviaa0}.

We have checked the relation \eqref{eq:F=ZSF} with the jet and soft functions given by \eqref{eq:lnZ_J} and \eqref{eq:deltaS}, respectively, by performing the two-loop calculation of $F(Q^2)$  and $\mathcal{F}(Q^2)$ in massless and massive QED. The massless form factor  $F(Q^2)$ was calculated with the conventions described in the first remark after Eq. \eqref{eq:massless_factorization_ggstar}. The massive form factor $F(Q^2, m^2)$  was calculated in the on-shell renormalization scheme, that is, the electron mass and wave function were renormalized.

Specializing Eq. \eqref{eq:massive_factorization} to the process of our present interest, we have
\begin{equation}
	{A}_\mu(e^+ e^-\rightarrow \gamma\gamma^*)=\left[Z_3^{\text{OS}}\right]^{1/2} Z_Je^{\delta S(-s-i0,  m^2)}\mathcal{A}_\mu(e^+ e^-\rightarrow \gamma\gamma^*)\,.
\end{equation}

%From the derivation of the soft function $S$ it is clear that for the process of present interest (\eeggs) it is obtained from \eqref{eq:lnS} by the substitution $Q^2\rightarrow -s -i0$. As for the jet function $Z_J$, it has to contain additional factor $Z_3$ from the photon wave function renormalization, so that the relation between  massive and massless amplitudes takes the form
%\be
%A(e^+ e^-\rightarrow \gamma\gamma^*)=Z_3 Z_J(m^2)\cdot S(-s-i0,  m^2)\cdot  A(e^+ e^-\rightarrow \gamma\gamma^*)~.  \label{A m = Z S A 0}
%\ee

%Using \eqref{eq:massless_factorization_ggstar} and \eqref{eq:massive_factorization}, we obtain the relation between
%\begin{equation}
%A_\mu(e^+ e^-\rightarrow \gamma\gamma^*)=e^R \cdot \mathcal{H}_\mu,   \label{A m = Z S A 0}
%\end{equation}
%where
%\be
%R = \ln\Big(Z_3 Z_J(m^2)\Big) +   \ln S(-s-i0,  m^2)+  \ln \mathcal{Z} . \label{R}
%\ee

\subsection{Soft virtual factorization in massive QED}
It is well known \cite{Yennie:1961ad} that in QED amplitude with massive electrons the singularities associated with soft virtual photons are factorized according to
\begin{equation}\label{eq:SoftVirtual}
	A(in\to out)=\exp\bigg\{-\sum_{i<j} Q_iQ_j V\left(p_i,p_j\right)\bigg\}H(in\to out)\,,
\end{equation}
where $H(in\to out)$, sometimes called ``hard amplitude'' in massive QED is finite and
%$e^{V^{(s)}}$, where

% where $e^{V^{(s)}_1}$ is the one-loop contribution. It means that there should be no stronger singularities in the exponent $R$ than $1/\e$; moreover, the singular terms must cancel in the sum $R+V^{(s)}_1$.
%It can be easily shown, writing, according to  \cite{Yennie:1961ad},   $V^{(s)}_1$ as
\begin{equation}
	V\left(p_i,p_j\right) = -\frac{e^2}{2}\int \tfrac{d^D k}{i(2\pi)^D}\tfrac{1}{k^2+i0}\Bigg(\tfrac{2p_i-k}{k^2-2(kp_i)+i0}+\tfrac{2p_j +k}{k^2+2(kp_j)+i0} \Bigg)^2\,,
\end{equation}
in dimensional regularization, cf. \cite[Eq. (2.23)]{Yennie:1961ad}. For our present goal we need the high-energy asymptotics of $V\left(p_+,p_-\right)$ which reads
\begin{equation}
V(p_+,p_-) =
%V
-a \left(\tfrac{m^2e^\gamma}{4\pi}\right)^{-\e}\left(-\tfrac{2 (\tLm-1)}{\e }+\tLm^2-\tLm-2\zeta_2+2+\mathcal{O}\left(\e ^1\right)\right)
%V/
\,,
\label{eq:V1}
\end{equation}
where  $a=\tfrac{\alpha}{4\pi}, \;\; \alpha = \tfrac{e^2}{4\pi}\approx 1/137.036$, $\tLm=\ln\tfrac{-s-i0}{m^2}$. Note that the representation \eqref{eq:SoftVirtual} is proven in \cite{Yennie:1961ad} using the on-mass shell renormalization scheme. Eq. \eqref{eq:SoftVirtual} then specifies as
\begin{equation}
	A_\mu(\eeggs)=e^{V(p_+,p_-)}H_\mu(\eeggs)
\end{equation}
On the other hand, the result of two previous subsections is the representation
\begin{equation}
	A_\mu(\eeggs)=\left(Z_3^{\text{OS}}\right)^{1/2}\,Z_J\, S\,\mathcal{Z}\,\mathcal{H}_\mu(\eeggs)
\end{equation}

Then the two finite quantities, $H_\mu(\eeggs)$, and $\mathcal{H}_\mu(\eeggs)$ are related via
\begin{equation}\label{eq:HviaHcal}
	H_\mu(\eeggs)=e^{-V(p_+,p_-)}\,\left(Z_3^{\text{OS}}\right)^{1/2}\,Z_J\, S\,\mathcal{Z}\,\mathcal{H}_\mu(\eeggs)\,,
\end{equation}
therefore the factor $e^{-V}\left(Z_3^{\text{OS}}\right)^{1/2}Z_JS\mathcal{Z}$ is also finite at $\e=0$. In terms of $\bar a$ its logarithm reads
\begin{multline}
	\ln\left[e^{-V}\left(Z_3^{\text{OS}}\right)^{1/2}Z_JS\mathcal{Z}\right]
	=\ln\left(e/e_{\MSb}\right)
	+\bar{a}[\tLmu^2-3 \tLmu+2 \tLm-\zeta _2+6]
	\\
	+\bar{a}^2\Big[
	\left(
	\tfrac{4 (\tLmu^3-\tLm^3)}{9}
	-\tfrac{38 \tLmu^2}{9}
	+\tfrac{8 \tLmu \tLm}{3}
	+\tfrac{14 \tLm^2}{9}
	+\tfrac{8 \zeta _2 (\tLmu-\tLm)}{3}
	+\tfrac{346 \tLmu}{27}
	-\tfrac{458 \tLm}{27}
	-\tfrac{4 \zeta _3}{9}
	-\tfrac{82 \zeta _2}{9}
	+\tfrac{5107}{162}
	\right)n_f
	\\
	+\tfrac{48\zeta _3-24 \zeta _2+3}{2} (\tLm-\tLmu)
	-42 \zeta _4
	-6 \zeta _3
	-48 \zeta _2 \ln{2}
	+22 \zeta _2
	+\tfrac{177}{8}
	\Big]
	+O(\bar {a}^3)\,,\label{eq:HviaHcal1}
\end{multline}
where $\tLmu=\ln\tfrac{-s-i0}{\mu^2}$, $\tLm=\ln\tfrac{-s-i0}{m^2}$
and we have separated the contribution \begin{multline}
	\ln\left(e/e_{\MSb}\right)=\tfrac12\ln\left({Z_3^{\text{OS}}}/{Z_3^{\MSb}}\right)=- \bar{a}\tfrac{2 n_f}{3}\left(\tLm-\tLmu\right)
	\\
	+\bar{a}^2\left[\tfrac{4}{9} n_f^2 \left(\tLm-\tLmu\right)^2- 2n_f\left(\tLm-\tLmu+\tfrac{15}{4}\right)\right]+\mathcal{O}(\bar{a}^3)\,,\label{eq:HviaHcal2}
\end{multline}
related to the difference between \MSb and on-shell charge.

\subsection{Factorization of soft radiation}

As it was already noticed in the previous section, the singularities in the  amplitude $A_\mu(\eeggs)$ come from infrared divergences. They disappear in the inclusive cross-section $d\sigma_{\textit{inc}}$ which accounts for soft photon emission. It is known, \cite{Bloch:1937pw, Yennie:1961ad}, that the cross section $d\sigma^{(n)}$ with emission of $n$ soft photons with energy less than $\omega_0$ each is given by the product
\begin{equation}
d\sigma^{(n)} = \tfrac{W^n(\omega_0)}{n!}d\sigma^{(0)}\,,
\end{equation}
where $d\sigma^{(0)}$ is the cross-section  of elastic (without soft emission) process   and $W(\omega_0)$ is the  probability of  soft photon emission  with energy less than $\omega_0$, so that the inclusive cross section $d\sigma_{\textit{inc}}(\omega_0)$ which includes emission  of soft photons with energy less than $\omega_0$
\begin{equation}
d\sigma_{\textit{inc}}(\omega_0) =\sum_{n=0}^\infty d\sigma^{(n)} = \exp[W(\omega_0)]d\sigma^{(0)}\,.  \label{eq:d_sigma_incl}
\end{equation}
In dimensional regularization we have the following form\footnote{Note that using the charge conservation condition $\sum Q_i=0$ we can identically rewrite Eqs. \eqref{eq:W} and \eqref{eq:Wint} as $W({in}\to {out}|\omega_0) = - e^2 \intop_{\omega<\omega_0}\tfrac{d^{D-1}k}{(2\pi)^{D-1}2\omega} \left(\sum_i\tfrac{Q_ip_i}{(k\cdot p_i) }\right)^2$.} of $W$:
\begin{equation}\label{eq:W}
	W(\omega_0) = W({in}\to {out}|\omega_0) =  -\sum_{i<j}Q_iQ_jW(p_i,p_j|\omega_0)\,,
\end{equation}
with
\begin{equation}
	W(p_i,p_j|\omega_0)=-e^2\intop_{\omega<\omega_0}\tfrac{d^{D-1}k}{(2\pi)^{D-1}2\omega}\left(\tfrac{p_i}{k\cdot p_i}-\tfrac{p_j}{k\cdot p_j}\right)^2 \label{eq:Wint}
\end{equation}
Using Eqs. (4.7)-(4.9) of Ref. \cite{Becher:2007cu}, we obtain the asymptotic form of $W(p_+,p_-|\omega_0)$:
\begin{multline}
W(p_+,p_-|\omega_0) =
%W
4a \tfrac{\Gamma(1-\e)}{ \Gamma(1-2\e)}\tfrac{(2\omega_0)^{-2\e}}{(4\pi)^{-\e}}\Big[-\e^{-1}\,(L-1) -\left(\tfrac12 (L-1)^2  +2\zeta_2 -\tfrac12\right)
\\
-\e\,\left(\tfrac16 (L-1)^3  +(2\zeta_2 -\tfrac12) L +2\zeta_3+\tfrac16\right) +\mathcal{O}\left(\e^2\right)\Big]
%W/
\,,\label{eq:W1}
\end{multline}
where $L=\ln\tfrac{s}{m^2}$ and the restriction $|\boldsymbol{k}|<\omega_0$ in \eqref{eq:W1} is imposed in the center-of-mass frame.
As we note below, the term of order $\e^1$ in the second line of this formula is irrelevant and falls out of the final result for inclusive cross section. Apart from this irrelevant term, Eq. \eqref{eq:W} can also be obtained by taking a large-$s$ asymptotics of the exact result in \cite[Eq. (41)]{Lee:2020zpo}.

\section{Consequences of factorization}\label{sec:consequences}
%The QED factorization properties derived in the previous section have several consequences.
Let us consider the factorization relation of the form
\begin{equation}\label{eq:fact}
	X(\e,z)=F(\e,z) Y(\e)\,,
\end{equation}
where $z$ is some small parameter and quantities $X$, $Y$ and $F$ admit perturbative expansions
\begin{equation}
	X(\e,z)=\sum_{l=0}^{\infty} X_l(\e,z) \tilde{a}^l\,,\qquad Y(\e)=\sum_{l=0}^{\infty} Y_l(\e) \tilde{a}^l\,,
	\qquad \ln F(\e,z) = \sum_{l=1}^{\infty} f_l(\e,\ln z) \tilde{a}^l\,
\end{equation}
with $f_l(\e,\ln z)$ being polynomial in $\ln z$. Here $\tilde{a}$ denotes a coupling constant in some renormalization scheme.

Suppose that $X(\e,z)$ is finite at $\e=0$.
Then we have two consequences of the factorization. First, in order to determine $X(0,z)$, we need to know factors $f_l(\e,\ln z)$ only up to $\e^0$ terms. Second, the terms in $X_L(0,z)$ amplified by powers of $\ln z$ are entirely expressed via lower-loop results $X_k(0,z)$, $k<l$.

The first statement immediately follows from Eq. \eqref{eq:fact} if we take logarithm of both left- and right-hand sides.
The second statement is also simple to understand. Indeed, $Y_l$ enters $X_l$ with unit coefficient (i.e., is not amplified by $\ln z$). Therefore, logarithmically amplified terms in $X_l$ depend on $Y_k$ with $k<l$. On the other hand, inverting Eq. \eqref{eq:fact} as
\begin{equation}
	Y(\e)=X(\e,z)/F(\e,z)\,,
\end{equation}
we can easily establish that $Y_l$ is expressed via $X_k$ with $k\leqslant l$. In particular, we have the relations
\begin{align}
	X_0(\e) &= Y_0(\e)\,,\nonumber\\
	X_1(\e,z) &= f_1(\e,\ln z) X_0(\e) + Y_1(\e)\,,\nonumber\\
	X_2(\e,z) &= \left[f_2(\e,\ln z)-\tfrac12f_1^2(\e,\ln z) \right] X_0(\e) + f_1(\e,\ln z) X_1(\e,z) + Y_2(\e)\,.
	\label{eq:X2viaX0X1Y2}
\end{align}
Note that this kind of reasoning is also valid when $z$ denotes a set of several small parameters.

Let us now specialize Eq. \eqref{eq:fact} to two cases. First, we put $z=\omega_0$, $X=d\sigma_{\textit{inc}}$, $Y=d\sigma_{0}(\omega_0)$, $F=e^{W(\omega_0)}$. Then Eq. \eqref{eq:fact} turns into Eq. \eqref{eq:d_sigma_incl}. Of course, the fact that $\ln \omega_0$-amplified terms in N$^n$LO approximation for $d\sigma_{\textit{inc}}$ are determined via N$^{k<n}$LO cross sections is well known. From the above considerations we conclude also that  $W(p_i,p_j|\omega_0)$ in Eqs. \eqref{eq:Wint} and \eqref{eq:W1} is sufficient to know only up to $\e^0$ terms.\footnote{Note that $W$ with this precision was calculated in Ref. \cite{Lee:2020zpo} exactly in $p_i\cdot p_j$ in arbitrary frame.} At first glance, the terms of order $\e^1$ in Eq. \eqref{eq:W1} are needed to keep terms of order $\e^0$ in the terms $\sigma^{(0)}_1 W$  and $\sigma^{(0)}_0 \tfrac12 W^2$, where $\sigma^{(0)}_B$   and $\sigma^{(0)}_1$ are  the elastic cross section in the Born  and the one-loop approximation, respectively.  However, it is easy to check that the order $\e$ part of  $W(\omega_0)$ cancels in these two contributions. We stress that such cancellation takes place not only in the two loop approximation considered here, but in all orders in $\alpha$.

Let us define the ``cross section'' of the process \eeggs as
\begin{equation}
	d\sigma^{A}_{\mu\nu} = {A}_{\mu\nu}(\eeggs) \tfrac{d\Phi}{2s}\,,
\end{equation}
where
\begin{equation}
	d\Phi=\tfrac{d^3 k}{(2\pi)^3 2\omega} (2\pi)\delta((p_++p_- -k)^2-q^2)
\end{equation}
and the leptonic tensor $A_{\mu\nu}(\eeggs)$, corresponding to the amplitude $A_{\mu\nu}(\eeggs)$ is defined as
\begin{equation}
	A_{\mu\nu}(\eeggs) = \overline\sum A_{\mu}(\eeggs)A^*_\nu(\eeggs)\,.
\end{equation}
Here $\overline\sum$ denotes the sum over the polarizations of the final photon and averaging over the polarizations of the initial electron and positron.
Similarly, $d\sigma^{H}$ and $d\sigma^{\mathcal{H}}$ are defined by the same formulae with $A$ replaced by $H$ and $\mathcal{H}$. The inclusive cross section is defined as
\begin{equation}
	d\sigma^{inc}_{\mu\nu} = {A}_{\mu\nu}^{inc}(\eeggs) \tfrac{d\Phi}{2s}\,,\label{eq:sigma_inc}
\end{equation}
where
$$
{A}_{\mu\nu}^{inc}(\eeggs)=e^{W(p_+,p_-|\omega_0)}\,{A}_{\mu\nu}(\eeggs)
$$
and $W(p_+,p_-|\omega_0)$ is defined in Eq. \eqref{eq:W1}.
Then we have the relation
\begin{equation}
	A_{\mu\nu}^{inc}=
	e^{W(p_+,p_-|\omega_0)}Z_3^{\text{OS}}\left|Z_JS\mathcal{Z}\right|^2
	\mathcal{H}_{\mu\nu}\,.\label{eq:AcalH}
\end{equation}
This relation is also a specialization of \eqref{eq:fact} with $z=\left\{\omega_0, m^2 \right\}$, $X=A^{inc}_{\mu\nu}$, $Y=\mathcal{H}_{\mu\nu}$, and $F=e^{W(p_+,p_-|\omega_0)}Z_3^{\text{OS}}\left|Z_JS\mathcal{Z}\right|^2$. As both $A^{inc}_{\mu\nu}$ and $\mathcal{H}_{\mu\nu}$ are finite at $\e=0$, we present the expression for $\ln F$ also at $\e=0$:
\begin{multline}
	\ln \left[e^{W(p_+,p_-|\omega_0)}Z_3^{\text{OS}}\left|Z_JS\mathcal{Z}\right|^2\right] = \ln\left(a/\aMS\right)
%logFih
	+2\aMS\,\left[
	\Lmu^2-3\Lmu+3\Lm
	-4 (\Lm-1)L_{\omega}
	-3\zeta _2
	+4\right]
	\\
	+\aMS^2\,\Big[
	\Big(
	\tfrac{8 (\Lmu^3- \Lm^3)}{9}
	-\tfrac{32}{3} (\Lmu-\Lm )(\Lm-1)L_{\omega}
	+\tfrac{4}{9} \left(L-L_{\mu }\right) \left(19 L_{\mu }+L\right)
	+\tfrac{548 \Lmu}{27}
	-\tfrac{772 \Lm}{27}
	-\tfrac{8 \zeta _3}{9}
	-\tfrac{164 \zeta _2}{9}
	\\
	+\tfrac{5107}{81}
	\Big)n_f
	-(48 \zeta _3 -24 \zeta _2 +3)(\Lmu-\Lm)
	+44 \zeta _2
	-12 \zeta _3
	-84 \zeta _4
	-96 \zeta _2 \ln{2}
	+\tfrac{177}{4}
	\Big]
%logFih/
	+\mathcal{O}\left(\aMS^3\right)\,,
	\label{eq:logFih}
\end{multline}
where $\Lmu=\Re\tLmu=\ln (s/\mu^2)$, $\Lm=\Re\tLm=\ln (s/m^2)$, $L_\omega=\ln\tfrac{\sqrt{s}}{2\omega_0}$.
We remind that this formula corresponds to the cross section \eqref{eq:d_sigma_incl} which includes radiation of one and two photons with the restriction $\omega < \omega_0$ imposed on energy of \emph{each} soft photon. If instead it is imposed on their total energy,  then in Eq. \eqref{eq:logFih} the additional term $-\aMS^232(L-1)^2\zeta_2$ appears. Let us note also, that the cross section \eqref{eq:d_sigma_incl} does not include the production of soft electron-positron pairs. It is precisely this circumstance that determines the presence of the term $-\aMS^2\tfrac{8 L^3}{9}$ in Eq. \eqref{eq:logFih}.

Note that in the context of measuring the hadronic contribution to the MAMM, what we really need is the cross section of the process in which the virtual photon goes into some hadronic state $h$. This cross section is expressed via ${A}_{\mu\nu}^{inc}(\eeggs)$ as
\begin{gather}
	d\sigma^{inc}(\eeggs\to \gamma h) = {A}_{\mu\nu}^{inc}(\eeggs) \tfrac{(4\pi \alpha)^2}{q^4|1-\mathcal{P}(q^2)|^2} J^{\mu}_h J^{*\nu}_h \tfrac{d\widetilde{\Phi}}{2s}\,,\nonumber\\
	d\widetilde{\Phi}=(2\pi)^4\delta(p_++p_- - k -q)\tfrac{d^3 k}{(2\pi)^3 2\omega} \prod_{f}\tfrac{d^3 p_f}{(2\pi)^3 2\varepsilon_f}\,.\label{eq:sigma_inc1}
\end{gather}
%
%\begin{equation}
%e^*_\mu(q)e_\nu(q) (2\pi)\delta((p_++p_- -k)^2-q^2)\rightarrow \tfrac{4\pi \alpha}{Q^4|1-\mathcal{P}(Q^2)|^2} J^{(f)}_\mu J^{*(f)}_\nu (2\pi)^4\delta(p_++p_- - k -P_f)d\rho_f~,
%\end{equation}
where $q=\sum_f p_f$,  $\mathcal{P}(q^2)$ is the polarization operator,   $J_h^\mu = \langle h|J^\mu(0)|0\rangle$ is the matrix element of the  hadronic current (not including the charge $e$)  between vacuum and the state $h$.  For example, in the case where the final state is $\pi_-(q_-)\pi_+(q_+)$ pair, we have $J_h^\mu J_h^{*\nu} = (q_- -q_+)^\mu (q_- -q_+)^\nu |F_\pi(q^2)|^2$, where $ F_\pi$ is the $\pi$-meson form factor. For the inclusive  production of hadrons this cross section is expressed via $R(q^2)$ ratio as
\begin{equation}
d\sigma^{inc}(\eeggs\to \gamma + \textit{hadrons}) =\tfrac{-g^{\mu\nu}}{2s} A_{\mu\nu}^{inc}\tfrac{d^3 k}{(2\pi)^3 2\omega}\tfrac{8\pi\alpha^2}{3q^2}R(q^2)\,.
\end{equation}
Note that the factor $\tfrac{1}{|1-\mathcal{P}(q^2)|^2}$ is included here in $R(q^2)$.

\section{Details of calculation}

As we have seen in two previous sections, the massive amplitude and the inclusive cross section of the process \eeggs are expressed in terms of hard amplitude $\mathcal{H}_\mu$ and the corresponding hard leptonic tensor $\mathcal{H}_{\mu\nu}$ at $\e=0$. We find these two objects to be the most convenient and fundamental and present our final results for them.

\begin{figure}
\begin{subfigure}{\textwidth}
	\includegraphics[width=\textwidth]{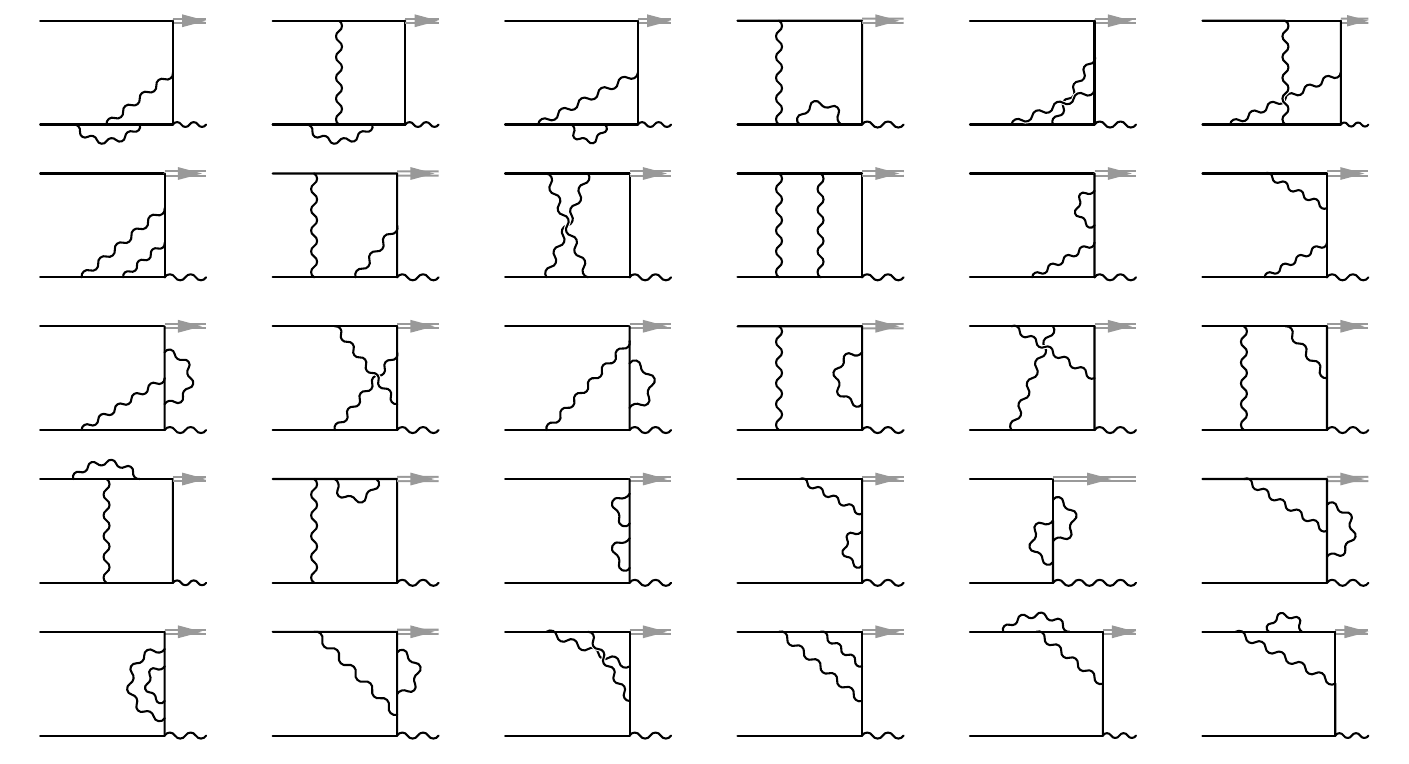}
	\caption{Pure photonic diagrams.}
	\label{fig:ph}
\end{subfigure}
\begin{subfigure}{\textwidth}
	\centering\includegraphics[width=0.66\textwidth]{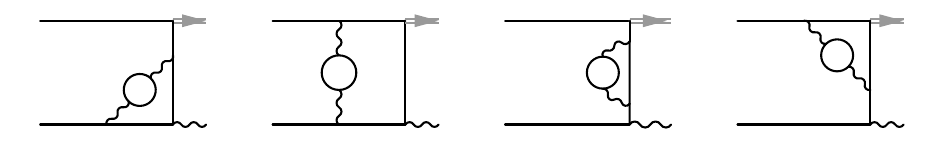}
	\caption{Diagrams with insertion of polarization operator.}
	\label{fig:po}
\end{subfigure}
\begin{subfigure}{\textwidth}
	\centering\includegraphics[width=0.66\textwidth]{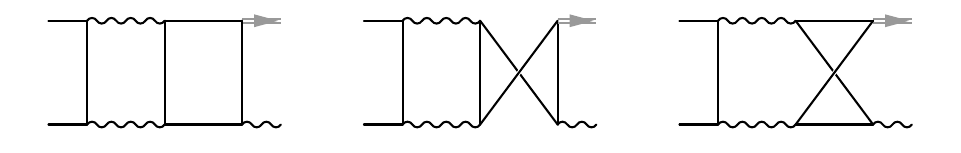}
	\caption{Diagrams with insertion of light-by-light block.}
	\label{fig:lbl}
\end{subfigure}
\caption{Two-loop diagrams for the process \eeggs.}
	\label{fig:diagrams}
\end{figure}

In order to derive the two-loop approximation for the hard amplitude, we have calculated the two-loop approximation for the massless amplitude $\mathcal{A}_\mu(\eeggs)$ and used the factorization formula \eqref{eq:massless_factorization_ggstar}.
The corresponding diagrams are shown in Fig. \ref{fig:diagrams}.

For the sake of brevity of the following formulas, it is convenient to define the dimensionless variables $T=-t/s$ and $U=-u/s$ and to put $s=1$. The full dependence of the result on $s$ can then be recovered from dimensional grounds. Here $t=(p_--k)^2$, $u=(p_+-k)^2$, $s=(p_-+p_+)^2$ are conventional Mandelstam variables.
The physical region is defined by inequalities
\begin{equation}
	T>0,\ U>0,\ q^2=1-T-U>0\,.
\end{equation}

\subsection{Tensor structure of amplitudes and leptonic tensors.}
The  amplitude $\mathcal{A}_\mu(\eeggs)$ can be represented as
\begin{equation}
	\mathcal{A}_\mu (\eeggs) =\sum_i \widehat{\mathcal{A}}_{n}\overline{u}(-p_+) \mathcal{T}_{\mu\sigma}^{(n)} u(p_-) e^{\sigma}(k)\,,
\end{equation}
where the tensors $\mathcal{T}_{\mu\sigma}^{(k)}$ are constructed of the external vectors and Dirac $\gamma$-matrices and the invariant amplitudes $\widehat{\mathcal{A}}_i$ are functions of $s,\, t,\, u$. In the same way we define invariant amplitudes $\widehat{\mathcal{H}}_n$ and $\widehat{H}_n$:
\begin{align}
	\mathcal{H}_\mu (\eeggs) &=\sum_i \widehat{\mathcal{H}}_{n}\overline{u}(-p_+) \mathcal{T}_{\mu\sigma}^{(n)} u(p_-) e^{*\sigma}(k)\,,\label{eq:calH}
	\\
	{H}_\mu (\eeggs) &=\sum_i \widehat{{H}}_{n}\overline{u}(-p_+) \mathcal{T}_{\mu\sigma}^{(n)} u(p_-) e^{*\sigma}(k)\,.
\end{align}

In $D$ dimensions there are $7$ linearly independent tensors which satisfy all requirements: $P$- and $C$-even, transverse to $q_\mu$ and $k_\sigma$ and not vanish upon bracketing with $\overline{u}(-p_+)\left[\ldots \right]u(p_-) e^{*\sigma}(k)$. They read
\begin{gather}
	\mathcal{T}^{(1)}_{\mu\sigma }=%\tfrac{1}{T}
	\left(\tfrac{p_{-\mu }}{1-T}-\tfrac{k_{\mu }}{T+U}\right)\left(\gamma _{\sigma }-\tfrac{2 \hat{k} p_{-\sigma }}{T}\right) \,,
	\quad
	\mathcal{T}^{(2)}_{\mu\sigma }=%\tfrac{1}{U}
	\left(\tfrac{p_{+\mu }}{1-U}-\tfrac{k_{\mu }}{T+U}\right) \left(\gamma _{\sigma }-\tfrac{2 \hat{k} p_{+\sigma }}{U}\right) \,,
	\nonumber\\
	\mathcal{T}^{(3)}_{\mu\sigma }= \left(\tfrac{p_{+\mu }}{1-U}-\tfrac{k_{\mu }}{T+U}\right) \left(\gamma _{\sigma }-\tfrac{2 \hat{k} p_{-\sigma }}{T}\right)\,,
	\quad
	\mathcal{T}^{(4)}_{\mu\sigma }=\left(\tfrac{p_{-\mu }}{1-T}-\tfrac{k_{\mu }}{T+U}\right) \left(\gamma _{\sigma }-\tfrac{2 \hat{k} p_{+\sigma }}{U}\right) \,,
	\nonumber\\
	\mathcal{T}^{(5)}_{\mu\sigma }=-\tfrac{p_{+\mu }}{1-U} \left(\gamma _{\sigma }-\tfrac{2 \hat{k} p_{-\sigma }}{T}\right)+\tfrac{p_{-\mu }}{1-T} \left(\gamma _{\sigma }-\tfrac{2 \hat{k} p_{+\sigma }}{U}\right)+\gamma _{\mu } \left(\tfrac{p_{-\sigma }}{T}-\tfrac{p_{+\sigma }}{U}\right) \,,
	\nonumber\\
	\mathcal{T}^{(6)}_{\mu\sigma }=\tfrac{1}{T U}\left(\hat{k} g_{\mu\sigma }-\gamma _{\sigma } k_{\mu }\right)+\tfrac{p_{+\mu } }{U(1-U)}\left(\gamma _{\sigma }-\tfrac{2 \hat{k} p_{-\sigma }}{T}\right)+\tfrac{p_{-\mu }}{T(1-T)} \left(\gamma _{\sigma }-\tfrac{2 \hat{k} p_{+\sigma }}{U}\right) \,,
	\nonumber\\
	\mathcal{T}^{(7)}_{\mu\sigma }=\tfrac{\gamma _{\sigma }\hat{k}\gamma _{\mu }-\gamma _{\mu }\hat{k}\gamma _{\sigma }}{2}
	+p_{+\mu } \left(T \gamma _{\sigma }-2 \hat{k} p_{-\sigma }\right)-p_{-\mu } \left(U \gamma _{\sigma }-2 \hat{k} p_{+\sigma }\right)+\gamma _{\mu } \left(U p_{-\sigma }-T p_{+\sigma }\right) \,.
	\label{eq:calT}
\end{gather}
On the other hand, by considering the $P$ and $C$ transformations of helicity amplitudes, we understand that in $d=4$ there are only $6$ independent structures are possible. Indeed, using Dirac equation for $\overline{u}(-p_+)$ and $u(p_-)$ it can be explicitly shown that
\begin{equation}
\overline{u}(-p_+)\mathcal{T}^{(7)}_{\mu\sigma }u(p_-) e^{*\sigma}(k)=\tfrac12\overline{u}(-p_+)\left[\gamma_{\perp\sigma}\hat{k}_\perp\gamma_{\perp\mu}-\gamma_{\perp\mu }\hat{k}_\perp\gamma_{\perp\sigma }\right]u(p_-) e^{*\sigma}(k)\,,
\end{equation}
where $v_\perp$ denotes the projection of $v$ onto the space orthogonal to  $(p_-,p_+)$ plane. At $d=4$ this space is two-dimensional and the quantity in square brackets vanishes identically.

The tensor structure of unpolarized leptonic tensors $A_{\mu\nu}=\overline{\sum}A_\mu A_{\nu}^*$,
$H_{\mu\nu}=\overline{\sum}H_\mu H_{\nu}^*$,
$\mathcal{H}_{\mu\nu}=\overline{\sum}\mathcal{H}_\mu \mathcal{H}_{\nu}^*$, and $A_{\mu\nu}^{inc}$ can be found in Ref. \cite{Rodrigo:2001kf}. For completeness, let us present the corresponding formula for, e.g, the hard leptonic tensor $\mathcal{H}_{\mu\nu}$:
\begin{multline}
	\mathcal{H}_{\mu\nu}=e_{\MSb}^2\big[g_{\mu\nu}^{\perp}\widehat{\mathcal{H}}_{00}
	+p_{-\mu}^\perp p_{-\nu}^\perp\widehat{\mathcal{H}}_{11}
	+p_{+\mu}^\perp p_{+\nu}^\perp\widehat{\mathcal{H}}_{22}
	+\left(p_{-\mu}^\perp p_{+\nu}^\perp+p_{+\mu}^\perp p_{-\nu}^\perp\right)\widehat{\mathcal{H}}_{12}\\
	+i\pi\left(p_{-\mu}^\perp p_{+\nu}^\perp-p_{+\mu}^\perp p_{-\nu}^\perp\right) \widehat{\mathcal{H}}_{-1}
	\big]\,,\label{eq:calHmn}
\end{multline}
where $g_{\mu\nu}^{\perp}=g_{\mu\nu}-q_\mu q_\nu/q^2$, $p_n^{\perp}=p_n-(p_n\cdot q)q/q^2$ and $\widehat{\mathcal{H}}_{mn}=\sum_{l\geqslant 0} \widehat{\mathcal{H}}_{mn}^{(l)}\aMS^l$. Note that the index $\perp$ can be omitted once we imply subsequent contraction with transverse hadronic tensor (in particular, it was done so in Ref. \cite{Rodrigo:2001kf}).

% Indeed, it can be explicitly shown that$\mathcal{T}^{(7)}_{\mu\sigma }$ identically vanish in $d=4$ when inserted between $\overline{u}(-p_+)\left[\ldots\right]u(p_-) e^{*\sigma}(k)$, 	since the space transverse to the $(p_1, p_2)$ plane is two-dimensional, so that $\gamma_\perp^{\sigma }\gamma_\perp^{\rho}\gamma_\perp^{\mu }-\gamma_\perp^{\mu }\gamma_\perp^{\rho}\gamma_\perp^{\sigma } =0$, and  all components of the  tensor $\mathcal{T}^{(7)}_{\mu\sigma }$, at least one of which lies in the $(p_1, p_2)$  plane, vanish due to the Dirac equation.

\subsection{IBP and DE reduction.}
Using the standard tensor decomposition approach, the invariant amplitudes $\mathcal{A}_k$ in the two-loop approximation can be written in terms of two-loop four-leg scalar master integrals first considered in Refs. \cite{GehrRem2001,GehrRem2001a} in the cross channel. The results of Refs. \cite{GehrRem2001,GehrRem2001a} are expressed in terms of products of harmonic polylogarithms of argument $s/q^2$ and Goncharov's polylogarithms of argument $t/q^2$ and with indices in the alphabet $\{0,1,-s/q^2,1-s/q^2\}$. In principle, the integrals which we need can be obtained by analytical continuation of the results of \cite{GehrRem2001,GehrRem2001a}, but we prefer to derive the expressions for the master integrals directly applicable to our process.\footnote{We have selectively checked that our results agree with those of Refs. \cite{GehrRem2001,GehrRem2001a}.} We use \texttt{LiteRed}  \cite{Lee2013a} and \texttt{Libra} \cite{Lee:2020zfb} to perform IBP reduction and reduction of differential equations, respectively. We find 85 master integrals in total among which there are 25 pairs related by the replacement $t\leftrightarrow u$. We reduce the differential equations with respect to $T$ and $U$ to $\e$-form. To fix the boundary constants we make the substitutions $T\to xT$ and  $U\to xU$ and consider the asymptotics $x\to 0$. The required terms of this asymptotics are calculated manually, using the expansion by regions technique and Mellin-Barnes parametrization.

As a result, we obtain the expressions for all required two-loop master integrals in terms of $\ln T,\ \ln U$ and Goncharov's polylogarithms $G(w_1,\ldots,w_n|1)$ with $w_k\in\{0,\tfrac1T,\tfrac1U,\tfrac1{T+U}\}$. The results for massless two-loop amplitude are expressed in terms of these functions with transcendental weight $n$ up to $4$.
We present our results for the master integrals in the ancillary files.

\section{Results}

Let us summarize the essential definitions for all quantities that we provide.
\begin{itemize}
	\item Hard massless amplitude and leptonic tensor:
	\begin{align}
		\mathcal{H}_\mu (\eeggs) &=e_{\overline{\text{MS}}}(\mu)\overline{u}(-p_+)\left[\sum_{n,l}\aMS^l\widehat{\mathcal{H}}_{n}^{(l)}  \mathcal{T}_{\mu\sigma}^{(n)} \right]u(p_-) e^{*\sigma}(k)\,,\label{eq:calHn.def}\\
		\mathcal{H}_{\mu\nu} (\eeggs) &=e_{\overline{\text{MS}}}^2(\mu)\sum_{mn\in I}
		\sum_{l}\aMS^l\widehat{\mathcal{H}}_{mn}^{(l)} {T}_{\mu\sigma}^{(mn)}\,,\label{eq:calHmn.def}
	\end{align}
	where $I=\{00,11,22,12,-1\}$,  $\mathcal{T}_{\mu\sigma}^{(n)}$ are defined in Eq. \eqref{eq:calT}, and
	\begin{multline}
	{T}_{\mu\sigma}^{(00)}=g_{\mu\nu}^{\perp},\
	{T}_{\mu\sigma}^{(11)}=p_{-\mu}^\perp p_{-\nu}^\perp,\
	{T}_{\mu\sigma}^{(22)}=p_{+\mu}^\perp p_{+\nu}^\perp,\
	{T}_{\mu\sigma}^{(12)}=
	p_{-\mu}^\perp p_{+\nu}^\perp+p_{+\mu}^\perp p_{-\nu}^\perp,\\
	{T}_{\mu\sigma}^{(-1)}=i\pi \left(p_{-\mu}^\perp p_{+\nu}^\perp-p_{+\mu}^\perp p_{-\nu}^\perp\right)\,,
	\end{multline}
	with $g_{\mu\nu}^{\perp}=g_{\mu\nu}-q_\mu q_\nu/q^2$, $p_n^{\perp}=p_n-(p_n\cdot q)q/q^2$.
	\item Hard massive amplitude and leptonic tensor:
	\begin{align}
		{H}_\mu (\eeggs) &=e\sum_{n,l} a^l\widehat{H}_{n}^{(l)}\overline{u}(-p_+) \mathcal{T}_{\mu\sigma}^{(n)} u(p_-) e^{*\sigma}(k)\,.\\
		{H}_{\mu\nu} (\eeggs) &=e^2\sum_{mn\in I}
		\sum_{l}a^l\widehat{H}_{mn}^{(l)} {T}_{\mu\sigma}^{(mn)}\,,
	\end{align}
	where ${T}_{\mu\sigma}^{(mn)}$ are defined above.
	\item Inclusive leptonic tensor
	\begin{equation}
	A^{inc}_{\mu\nu} (\eeggs) =e^2\sum_{mn\in I}
	\sum_{l}a^l\widehat{A}^{inc(l)}_{mn} {T}_{\mu\sigma}^{(mn)}\,,\label{eq:Ainc}
	\end{equation}
\end{itemize}
Note that the above quantities are finite at $\e=0$, in contrast to the massless and massive amplitudes $\mathcal{A}_\mu$ and ${A}_\mu$. The hard massive and massless amplitudes, ${H}_\mu (\eeggs)$ and $\mathcal{H}_\mu (\eeggs)$, are related via Eqs. \eqref{eq:HviaHcal}--\eqref{eq:HviaHcal2}. The inclusive leptonic tensor $A^{inc}_{\mu\nu} (\eeggs)$ and hard massless leptonic tensor $\mathcal{H}_{\mu\nu} (\eeggs)$ are related via Eqs. \eqref{eq:AcalH} and \eqref{eq:logFih}. Let us also note the relation
\begin{equation}
	A^{inc}_{\mu\nu}(\eeggs)=\exp\left[W(p_+,p_-|\omega_0)+2\Re V(p_+,p_-)\right]H_{\mu\nu} (\eeggs)\,.
\end{equation}
We stress that we define perturbative expansion of $\mathcal{H}_\mu$ and $\mathcal{H}_{\mu\nu}$ in terms of \MSb coupling constant $\aMS=\frac{\alpha_{\MSb}(\mu)}{4\pi}=\frac{e_{\MSb}^2(\mu)}{(4\pi)^2}$, while that of ${H}_\mu$, ${H}_{\mu\nu}$, and $A^{inc}_{\mu\nu}$ in terms of on-shell coupling constant $a=\frac{\alpha}{4\pi}=\frac{e^2}{(4\pi)^2}\approx(4\pi)^{-1}/137.036$.
Below we present our results.

\subsection{Results for hard massless invariant amplitudes}
For the hard invariant amplitudes $\widehat{\mathcal{H}}_n^{(l)}$ defined in Eq. \eqref{eq:calHn.def} we  have
\begin{equation}
	\left\{\widehat{\mathcal{H}}_1^{(0)},\ldots, \widehat{\mathcal{H}}_7^{(0)}\right\}=\left\{0,\,0,\,0,\,0,\,2-T-U,\,U-T,\,\tfrac{T+U}{T U}\right\}\,,
\end{equation}
The one-loop results read
\begin{multline}
	\left\{\widehat{\mathcal{H}}_1^{(1)},\ldots, \widehat{\mathcal{H}}_7^{(1)}\right\}
	=\big\{
%hcH1
	-2 \left(F_1-F_3\right) (1-T)/T,\,
%hcH1/
%hcH2
	2 \left(F_2-F_4\right) (1-U)/U,\,
%hcH2/
%hcH3
	2 F_3,\,
%hcH3/
%hcH4
	-2 F_4,\,
%hcH4/
%hcH5
	\\
	(1-T) F_6 + (1-U) F_5 + F_3 + F_4,\,
%hcH5/
%hcH6
	(1-T) F_6 - (1-U) F_5 -F_3 + F_4,\,
%hcH6/
%hcH7
	F_7
%hcH7/
	\big\}\,,
\end{multline}
where the functions $F_k$ have the form
\begin{align}
%F1
	F_1&=\tfrac{2 T-1}{T-1}+\tfrac{2 q^2 \ln \left(q^2\right)}{T+U}-\tfrac{q^2 T \ln \left(-\tfrac{U}{q^2}\right)}{(T-1)^2}
%F1/
	\,,\nonumber\\
%F2
	F_2&=\tfrac{2 U-1}{U-1}+\tfrac{2 q^2 \ln \left(q^2\right)}{T+U}-\tfrac{q^2 U \ln \left(-\tfrac{T}{q^2}\right)}{(U-1)^2}
%F2/
	\,,\nonumber\\
%F3
	F_3&=\tfrac{(T+U) f(U,T)}{T}+\tfrac{2 \ln \left(q^2\right)}{T+U}-\tfrac{(T+2 U) \ln \left(-\tfrac{U}{q^2}\right)}{T-1}+1
%F3/
	\,,\nonumber\\
%F4
	F_4&=\tfrac{(T+U) f(T,U)}{U}+\tfrac{2 \ln \left(q^2\right)}{T+U}-\tfrac{(2 T+U) \ln \left(-\tfrac{T}{q^2}\right)}{U-1}+1
%F4/
	\,,\nonumber\\
%F5
	F_5&=f(T,U)-g\left(\mu ^2\right)-\tfrac{2 \ln \left(q^2\right)}{T+U}+\tfrac{T \ln \left(-\tfrac{U}{q^2}\right)}{T-1}
%F5/
	\,,\nonumber\\
%F6
	F_6&=f(U,T)-g\left(\mu ^2\right)-\tfrac{2 \ln \left(q^2\right)}{T+U}+\tfrac{U \ln \left(-\tfrac{T}{q^2}\right)}{U-1}
%F6/
	\,,\nonumber\\
%F7
	F_7&=\tfrac{f(T,U)}{T}+\tfrac{f(U,T)}{U}-\tfrac{(T+U) g\left(\mu ^2\right)}{T\, U}+\tfrac{3 \ln \left(-\tfrac{T}{q^2}\right)}{1-U}+\tfrac{3 \ln \left(-\tfrac{U}{q^2}\right)}{1-T}
%F7/
	\,.
\end{align}
We remind that $q^2=1-T-U$.
Here
\begin{align}
%fdef
	f(T,U)&=2 \mathrm{Li}_2(T+U)-2 \mathrm{Li}_2\left(\tfrac{T}{1-U}\right)+2 \ln (-T) \ln (1-U)-\ln^2(1-U)
%fdef/
	\,,\nonumber\\
%gdef
	g(\mu^2)&=\ln ^2\left(-\mu ^2\right)+3 \ln\left(-\mu ^2/q^2\right)-\zeta_2+9
%gdef/
	\,.\label{eq:fg}
\end{align}
In the above formulae we use the convention $\ln(-x)=\ln x+i\pi$ for $x>0$.
The two-loop hard invariant amplitudes $\widehat{\mathcal{H}}_1^{(2)},\ldots, \widehat{\mathcal{H}}_7^{(2)}$ can be found in ancillary files.

\subsection{Results for hard massless leptonic tensor}
We have
\begin{equation}
	\left\{\widehat{\mathcal{H}}_{00}^{(0)},\widehat{\mathcal{H}}_{11}^{(0)},\widehat{\mathcal{H}}_{22}^{(0)},\widehat{\mathcal{H}}_{12}^{(0)},\widehat{\mathcal{H}}_{-1}^{(0)}\right\}=
	\left\{-\tfrac{q^2+T^2+U^2}{T U},-\tfrac{4q^2}{TU},-\tfrac{4q^2}{TU},0,0\right\}\,.
\end{equation}
In the one-loop approximation  we have
\begin{align}
\widehat{\mathcal{H}}_{00}^{(1)}=&
%	\left(6 \ln {\tfrac{q^2}{\mu^2}}-2 \Lmu^2+14 \zeta_2-18\right)
%H00
	-2\Re g(\mu^2)\widehat{\mathcal{H}}_{00}^{(0)}
	+4 \tfrac{T^2-T+U^2-U}{T\, U\, (1-q^2)}\ln{q^2}
	-\tfrac{2 (1-U) (2+T)}{(1-T) T}\ln \left(\tfrac{U}{q^2}\right)
	-\tfrac{2 (1-T) (2+U)}{(1-U) U} \ln\left(\tfrac{T}{q^2}\right)
	\nonumber\\&
	-2 \tfrac{1+q^2}{T\, U}
	-2\left(\tfrac{q^2}{U^2}+\tfrac{2-2 U+U^2}{T\, U}\right)\Re f(T,U)
	-2\left(\tfrac{q^2}{T^2}+\tfrac{2-2 T+T^2}{T\, U}\right)\Re f(U,T)
%H00/
	\,,
	\\
\widehat{\mathcal{H}}_{11}^{(1)}=&
%	\left(6 \ln {\tfrac{q^2}{\mu^2}}-2 \Lmu^2+14 \zeta_2-18\right)
%H11
	-2\Re g(\mu^2)\widehat{\mathcal{H}}_{11}^{(0)}
	+\tfrac{8 q^2 }{U^2}\left(\tfrac{2 \left(1-U+U^2\right)}{\left(1-q^2\right)^2}-\tfrac{2 (1-U)}{\left(1-q^2\right) T}-\tfrac{(1-U)^2}{T^2}\right)\ln{q^2}
	\nonumber\\&
	+\tfrac{4 (1-U) (1+q^2)}{(1-T) T\, U}
	-\tfrac{16 (1-U-T\, U)}{T\, U\, (1-q^2)}
	-\tfrac{4 q^2\, (2+U)}{T\, U^2}\ln \left(\tfrac{T}{q^2}\right)
	-\tfrac{4 q^2\, (2-3 T) (1-U)^2}{(1-T)^2 T^2 U}\ln \left(\tfrac{U}{q^2}\right)
	\nonumber\\&
	-4 q^2 \tfrac{1+U^2}{T\, U^3}\Re f(T,U)
	-4 q^2\tfrac{ 2 T^2+(1-U)^2}{T^3 U}\Re f(U,T)
%H11/
	\,,
	\\
\widehat{\mathcal{H}}_{22}^{(1)}=&
	\widehat{\mathcal{H}}_{11}^{(1)}(T\leftrightarrow U)\,,
	\\
\widehat{\mathcal{H}}_{12}^{(1)}=&
%H12
	\tfrac{24}{T\, U}-\tfrac{16 (1-T\, U)}{T\, U\, (T+U)}-\tfrac{4 (2-T)}{(1-T) T}-\tfrac{4 (2-U)}{(1-U) U}
	-\tfrac{8q^2}{T\, U}\left(\tfrac{q^2}{T\, U}-\tfrac{1}{T+U}+\tfrac{2 (1-T\, U)}{(T+U)^2}\right)\ln{q^2}
	\nonumber\\&
	-\tfrac{4q^2}{(1-U)^2} \left(2+\tfrac{(2-U)q^2}{T\, U^2}\right)\ln \left(\tfrac{T}{q^2}\right)
	-\tfrac{4q^2}{(1-T)^2} \left(2+\tfrac{(2-T)q^2}{U T^2}\right)\ln \left(\tfrac{U}{q^2}\right)
	\nonumber\\&
	-\tfrac{4q^2\,(1-T-T\, U)}{TU^3}\Re f(T,U)
	-\tfrac{4q^2\,(1-U-T\, U)}{UT^3}\Re f(U,T)
%H12/
	\,,
	\\
\widehat{\mathcal{H}}_{-1}^{(1)}=&
%Hm1
	\tfrac{4q^2\,(T-U)\left(T^2 U+T\, U^2+2q^2\right)}{(1-T)^2 T\, (1-U)^2 U}+\tfrac{8 q^2}{T^2 U}\ln (1-T)-\tfrac{8q^2}{T\, U^2}\ln (1-U)
%Hm1/
	\,.
\end{align}
The functions $f(T,U)$ and $g(\mu^2)$ are defined in Eq. \eqref{eq:fg}, so that
\begin{align*}
	\Re f(T,U)&=2 \mathrm{Li}_2(T+U)-2 \mathrm{Li}_2\left(\tfrac{T}{1-U}\right)+2 \ln{T} \ln (1-U)-\ln^2(1-U)\,,\nonumber\\
	\Re g(\mu^2)&=\ln ^2\left(\mu ^2\right)+3 \ln\left(\mu ^2/q^2\right)-7\zeta_2+9\,.
\end{align*}

The two-loop results $\widehat{\mathcal{H}}_{00}^{(2)},\ \widehat{\mathcal{H}}_{11}^{(2)},\ \widehat{\mathcal{H}}_{22}^{(2)},\ \widehat{\mathcal{H}}_{12}^{(2)},\ \widehat{\mathcal{H}}_{-1}^{(2)}$ can be found in ancillary files.

\subsection{Results for inclusive cross section}
For reader convenience we present in the ancillary files our results for the coefficients ${\widehat{A}}_{mn}^{inc(l)}$  defined in Eq. \eqref{eq:Ainc}.
We have checked that our results for ${\widehat{A}}_{mn}^{inc(0)}$ (Born approximation) and ${\widehat{A}}_{mn}^{inc(1)}$ (one-loop approximation) agree with those of Refs. \cite{Rodrigo:2001kf,kuhn2002radiative}.\footnote{The first paper \cite{Rodrigo:2001kf} contains incorrect expression for $\widehat{A}^{inc\,(1)}_{-1}$. There are also typos in \cite[Eq. (16)]{kuhn2002radiative}, which forced us to invert Eq. (15) of that paper instead.}

The general form of the two-loop contributions $\widehat{A}_{mn}^{inc(2)}$ is the following
\begin{equation}\label{eq:Anminc}
	\widehat{A}_{mn}^{inc(2)}=c_0+c_1 L+c_2 L^2+c_3 L^3+c_4 L_{\omega }+c_5 L L_{\omega }+c_6 L^2 L_{\omega }+c_7 L_{\omega }^2+c_8 L L_{\omega }^2+c_9 L^2 L_{\omega }^2\,,
\end{equation}
where the coefficients $c_{0-9}$ are some functions of $T$ and $U$ and $L_\omega=\ln\frac{\sqrt{s}}{2\omega_0}$.
As we explained in the introduction, there are contributions to the cross section $d\sigma(e^+e^-\to \gamma f)$ of the relative magnitude $(\alpha/\pi)^2$, which can not be calculated \textit{ab initio}. The contribution of the term $c_0$ is of the same order. Remarkably, other terms in Eq. \eqref{eq:Anminc}, amplified by large logarithms, can be expressed via  $\widehat{A}_{mn}^{inc(0)}$ and $\widehat{A}_{mn}^{inc(1)}$, as we explained in Section \ref{sec:consequences}. The specification of last relation in \eqref{eq:X2viaX0X1Y2} to the case $X=\widehat{A}_{mn}$, $Y=\widehat{\mathcal{H}}_{mn}$ reads
\begin{multline}
%hiA
\widehat{A}_{mn}^{inc(2)} =
\Big[-32 (L-1)^2 \left(L_{\omega }-\tfrac{3}{4}\right)^2+16\left(7-3\zeta _2\right) (L-1) \left(L_{\omega }-\tfrac{3}{4}\right)
\\
+\left(48 \zeta _3-24 \zeta _2+3\right) L-129 \zeta _4-12 \zeta _3-96\zeta _2 \ln{2}+128\zeta _2-\tfrac{215}{4}
\\
+n_f\, \left(\tfrac{32}{3} (L-1) L\, \left(L_{\omega }-\tfrac{3}{4}\right)-\tfrac{8 L^3}{9}+\tfrac{76 L^2}{9}-\tfrac{988 L}{27}-\tfrac{8 \zeta _3}{9}-\tfrac{164 \zeta _2}{9}+\tfrac{5107}{81}\right)
\Big]\widehat{A}_{mn}^{inc(0)}\\
+\left[-8 (L-1) \left(L_{\omega }-\tfrac{3}{4}\right)-6 \zeta _2+14+\tfrac{4}{3} n_f L\right]\widehat{A}_{mn}^{inc(1)}
+\left.\widehat{\mathcal{H}}_{mn}^{(2)}\right|_{\mu^2=s}
%hiA/
\end{multline}
The last term $\left.\widehat{\mathcal{H}}_{mn}^{(2)}\right|_{\mu^2=s}$ corresponds to the two-loop hard leptonic tensor coefficients taken at $\mu^2=s$ and does not contain large logarithms.

\subsection{Small-$q^2$ asymptotics}
It is interesting to consider the small-$q^2$ asymptotics of the obtained formulae. This asymptotics is important for the experiments on measuring $R(s)$ in the region of relatively small energy $\sqrt{s}$ by radiative return from the high energy of the collider. Also this asymptotic allows one to obtain the cross section of the process $e^+e^-\to \gamma\gamma$. Comparison of the result  obtained in this way with the expected result of a direct calculation of this cross section would be a rigorous check of the factorization formulas used here.

On general ground, one would expect the appearance of terms amplified by $\ln q^2$. Indeed, at first glance the asymptotics of the one-loop hard amplitudes does contain $\ln{q^2}$:
\begin{align}
	\label{eq:Hcal1.asy}
	\widehat{\mathcal{H}}_1^{(1)}&\approx
	-2\ln{q^2}
	+\tfrac{2 U}{T^2}\left[\widetilde{L}_U^2+6\zeta_2\right]+
	\tfrac{2 (1+U)}{T} \widetilde{L}_U+2,
	\\
	\widehat{\mathcal{H}}_2^{(1)}&\approx
	2\ln {q^2}
	-\tfrac{2 T }{U^2}\left[\widetilde{L}_T^2+6\zeta_2\right]-\tfrac{2 (1+T)}{U} \widetilde{L}_T-2,
	\\
	\widehat{\mathcal{H}}_3^{(1)}&\approx
	-\tfrac{2 T \ln {q^2}}{U}+\tfrac{2 \widetilde{L}_U^2}{T}+\tfrac{2 (U+1) \widetilde{L}_U}{U}+\tfrac{12 \zeta _2}{T}+2,
	\\
	\widehat{\mathcal{H}}_4^{(1)}&\approx\tfrac{2 U \ln {q^2}}{T}-\tfrac{2 \widetilde{L}_T^2}{U}-\tfrac{2 (T+1) \widetilde{L}_T}{T}-\tfrac{12 \zeta _2}{U}-2,
	\\
	\widehat{\mathcal{H}}_5^{(1)}&\approx
	\left(1+T U\right)\left[\tfrac{\widetilde{L}_T^2}{U}+\tfrac{\widetilde{L}_U^2}{T}+\tfrac{6\zeta _2}{TU}\right] +(3-T) \widetilde{L}_T+(3-U) \widetilde{L}_U-g(\mu^2)+2,
	\\
	\widehat{\mathcal{H}}_6^{(1)}&\approx\left(1-TU\right) \left[\tfrac{\widetilde{L}_T^2}{U}-\tfrac{\widetilde{L}_U^2}{T}+\tfrac{ (T-U)\zeta _2}{T U}\right]+(3-T) \widetilde{L}_T-(3-U) \widetilde{L}_U
	+(T-U)g(\mu^2),
	\\
	\widehat{\mathcal{H}}_7^{(1)}&\approx\tfrac{\widetilde{L}_T^2}{T}+\tfrac{\widetilde{L}_U^2}{U}+\tfrac{3 \widetilde{L}_T}{T}+\tfrac{3 \widetilde{L}_U}{U}-\tfrac{g(\mu^2)}{T U}+\tfrac{6 \zeta _2}{T U},
	\label{eq:Hcal7.asy}
\end{align}
where $\widetilde{L}_T=\ln(-T)=\ln T+i\pi$, $\widetilde{L}_U=\ln(-U)=\ln U+i\pi$ and we imply that $U=1-T$. We see that the first four invariant amplitudes do contain $\ln q^2$. However, we have to take into account that in the limit $q^2\to 0$ the following relations hold:
\begin{align}
	U\mathcal{T}^{(1)}_{\mu\sigma }+T\mathcal{T}^{(3)}_{\mu\sigma }&\approx q_\mu\left(T\gamma _{\sigma }-2 \hat{k} p_{-\sigma }\right)\,,
	\\
	T\mathcal{T}^{(2)}_{\mu\sigma }+U\mathcal{T}^{(4)}_{\mu\sigma }&\approx q_\mu\left(U\gamma _{\sigma }-2 \hat{k} p_{+\sigma }\right)\,.
\end{align}
Therefore, when $\mathcal{H}_{\mu}$ is multiplied by transverse hadronic current $J_h^\mu$, such that $J_h\cdot q=0$, only the combinations $T\widehat{\mathcal{H}}_1-U\widehat{\mathcal{H}}_3$ and $U\widehat{\mathcal{H}}_2-T\widehat{\mathcal{H}}_4$ of the first four invariant amplitudes enter the asymptotics. It is easy to see from Eqs. \eqref{eq:Hcal1.asy}--\eqref{eq:Hcal7.asy} that the corresponding combinations of the one-loop invariant amplitudes do not contain $\ln q^2$. We have checked that the same is true for the two-loop invariant amplitudes $\widehat{\mathcal{H}}_n^{(2)}$.

Let us present also the corresponding expressions for the coefficients $\widehat{\mathcal{H}}_{mn}^{(1)}$ entering hard leptonic tensor:
\begin{gather}
	\label{eq:Hcal00.asy}
	\widehat{\mathcal{H}}_{00}^{(1)}\approx
	-\tfrac{2 \left(1+T^2\right) L_T^2}{T U}-\tfrac{2 \left(1+U^2\right) L_U^2}{T U}
	-\tfrac{2 (3-T) L_T}{T}-\tfrac{2 (3-U) L_U}{U}
	+ (1-2 T U)\tfrac{2 \Re g(\mu^2)}{T U}-\tfrac{2}{T U}\,,
	\nonumber\\
	\widehat{\mathcal{H}}_{11}^{(1)}\approx \tfrac{4 (T-U)^2}{U^2}\,,\qquad
	\widehat{\mathcal{H}}_{22}^{(1)}\approx \tfrac{4 (T-U)^2}{T^2}\,,\qquad
	\widehat{\mathcal{H}}_{12}^{(1)}\approx -\tfrac{4 (T-U)^2}{TU}\,,\qquad
	\widehat{\mathcal{H}}_{-1}^{(1)}\approx 0\,,
\end{gather}
where ${L}_T=\ln T$, ${L}_U=\ln U$ and we imply that $U=1-T$.
We see that $\widehat{\mathcal{H}}_{mn}^{(1)}$ do not contain $\ln q^2$. The same is true for the two-loop coefficients $\widehat{\mathcal{H}}_{mn}^{(2)}$.

The asymptotic expressions for $\widehat{\mathcal{H}}_{n}^{(2)}$ and $\widehat{\mathcal{H}}_{mn}^{(2)}$ can be found in ancillary files.

\subsection{Small-angle asymptotics}

Another region important for the applications to measuring $R(s)$ is $T\ll 1$. This region corresponds to small angles of the radiated photon. The differential cross section in this region scales as $(\ln T)^l/T$, where $l$ is the number of loops.

For one-loop coefficients $\widehat{\mathcal{H}}_{mn}^{(1)}$ we have:
\begin{align}
	\label{eq:Hcal00.Tasy}
	\widehat{\mathcal{H}}_{00}^{(1)}&\approx
	\tfrac1T\Big\{
	2\tfrac{1+q^4}{U}\left[2 \mathrm{H}_1  L_T-2 \mathrm{H}_2
	+\mathrm{H}_1^2 +L_{\mu }^2+3 \mathrm{H}_1
	-3 L_{\mu }- 7\zeta _2+9\right] -\tfrac{2 \left(2-U\right)}{U}
	\Big\}
	,
	\\
	\widehat{\mathcal{H}}_{11}^{(1)}&\approx
	\tfrac1T\Big\{\tfrac{4(1-U) \left(1+U^2\right)}{U^3}\left[2\mathrm{H}_1 L_T
	-2\mathrm{H}_2+\mathrm{H}_1^2\right]
	+\tfrac{8 (1-U) }{U}\left[L_{\mu }^2+3 \mathrm{H}_1-3L_{\mu }-7\zeta _2+9\right]
	\nonumber\\&
	-\tfrac{4 (1-U) (2+U)}{U^2} L_T
	+\tfrac{4 (1-U) \left(5-8 U\right)}{U^3} \mathrm{H}_1
	-\tfrac{12 (1-U)}{U^2}
	\Big\}
	,
	\\
	\widehat{\mathcal{H}}_{22}^{(1)}&\approx
	\tfrac1T\Big\{
	\tfrac{4 (1-U) \left(1+2 U^2\right) }{U^3}\left[2 \mathrm{H}_1 L_T-2 \mathrm{H}_2+\mathrm{H}_1^2\right]+\tfrac{8 (1-U)}{U} \left[L_{\mu }^2+3 \mathrm{H}_1-3 L_{\mu }-7 \zeta _2+9\right]
	\nonumber\\&
	-\tfrac{4 (2-3 U)}{(1-U) U^2} L_T
	+\tfrac{20-16U(2-U)^2}{(1-U) U^3}\mathrm{H}_1
	-\tfrac{4 (3-4 U)}{(1-U) U^2}
	\Big\}
	,
	\\
	\widehat{\mathcal{H}}_{12}^{(1)}&\approx
	\tfrac1T\Big\{
	\tfrac{4 (1-U)}{U^3} \left[2 \mathrm{H}_1 L_T+\mathrm{H}_1^2-2 \mathrm{H}_2\right]
	-\tfrac{4 (2-U)}{U^2} L_T
	-\tfrac{4 \left(2 U^3-8 U^2+12 U-5\right)}{U^3}\mathrm{H}_1
	-\tfrac{4 (3-U) (1-U)}{U^2}
	\Big\}
	,
	\\
	\widehat{\mathcal{H}}_{-1}^{(1)}&\approx
	\tfrac1T\Big\{
	\tfrac{8  (1-U)}{U^2}\mathrm{H}_1-\tfrac{8}{U}
	\Big\}
	\,.
\end{align}
Here $\mathrm{H}_n=\mathrm{H}_n(U)$ are harmonic polylogarithms \cite{remiddi2000harmonic}. In particular, we have
\begin{equation*}
	\mathrm{H}_1(U)=\text{Li}_1(U)=-\ln(1-U),\quad\mathrm{H}_2(U)=\mathrm{Li}_2(U)\,.
\end{equation*}
The asymptotics of the two-loop coefficients has a form $\widehat{\mathcal{H}}_{mn}^{(2)}=T^{-1}p_{mn}(L_T)$, where $p_{mn}(x)$ are the second-order polylomials with coefficients expressed via
\begin{align*}
	\mathrm{H}_n(U)&=\mathrm{Li}_n(U) \text{ with }n=1,2,3,4\,,
	\\
	\mathrm{H}_{2,1}(U)&=\text{S}_{1,2}(U)=-\text{Li}_3(1-U)+\zeta_3+\text{Li}_2(1-U) \ln (1-U)+\tfrac{1}{2} \ln{U} \ln ^2(1-U)\,,
	\\
	\mathrm{H}_{3,1}(U)&=\text{S}_{2,2}(U)\,.
\end{align*}
Here $\mathrm{S}_{n,m}(U)$ denotes the Nielsen polylogarithm. The asymptotic expressions for the invariant amplitudes $\widehat{\mathcal{H}}_{n}^{(l)}$ and for the two-loop coefficients $\widehat{\mathcal{H}}_{mn}^{(2)}$ can be found in the ancillary files.

\subsection{Soft-photon asymptotics}

It is well known that the amplitudes and cross sections of processes with soft photon emission are given by the  products of the accompanying radiation factors and the amplitudes and cross sections of processes without radiation.  According to this,  when the real photon is soft, we should have
\begin{equation}
	{A}_\mu (\eeggs) = e \left(\tfrac{p_-}{(k\cdot p_-)}-\tfrac{p_+}{(k \cdot  p_+)}\right)^\nu e_{\nu}(k) F(s)\gamma_\mu +\mathcal{O}(\omega^0)\,.\label{eq:soft}
\end{equation}

An interesting question is the region of applicability of this relation. For the cross-channel process $e(p_1)+\gamma^*(q) \rightarrow e(p_2) +\gamma(k)$ with $q^2<0$, $Q^2 = - q^2\gg m^2 $  this question was discussed in the one-loop approximation in   \cite{Kuraev1987}.  It was shown there that this region is defined by the conditions
\begin{equation}
(kp_-)\ll s~, \quad (kp_+)\ll s\,,\label{eq:softregion}
\end{equation}
that is, it is much wider than not only the region restricted by the conditions
\begin{equation}
(kp_-)\ll m^2~, \;\; (kp_+)\ll m^2~,
\end{equation}
which could be expected based on the requirements for the small departure of the radiating particle from the mass surface, but also a much wider region
\begin{equation}
\frac{(kp_-)(kp_+)}{s}\ll m^2~,
\end{equation}
in which the applicability of the accompanying bremsstrahlung factor in hadron physics was shown by  V.N. Gribov \cite{Gribov:1966hs}.

Since  in Eq. \eqref{eq:soft} both ${A}_\mu (\eeggs)$ and $F(s)$ are IR divergent, it is convenient to compare the finite ``hard massive'' amplitudes $H_\mu (\eeggs)=e^V{A}_\mu (\eeggs)$ and $e^V F(s)$, where $V$ is defined in Eq. \eqref{eq:V1}.
We have checked that the relation \eqref{eq:soft}, when multiplied by $e^V$, indeed holds provided the conditions \eqref{eq:softregion} hold. Thus, we have checked that the statement about a wide range of applicability remains valid in two loops.
\subsection{Description of ancillary files.}

As many of our results are too lengthy to present them in the text, we provide the following ancillary files:
\begin{itemize}
	\item \texttt{MIs/}
	\begin{itemize}
	\item \texttt{MIs.def} --- definition of master integrals in terms of integrand in the momentum space.
	\item \texttt{MIs.graphs.pdf} --- a picture of all master integrals (blue, green, black, and red external leg corresponds to incoming $p_-,\ p_+,\ -k$, and $-q$ momentum, respectively).
	\item \texttt{MIs2UTs.rules} --- expression of master integrals via uniform transcendental basis (canonical master integrals).
	\item \texttt{UTs2Gs.rules} --- expression of uniform transcendental basis via Goncharov's polylogarithms $G$.
	\end{itemize}
	\item \texttt{calHn/}
	\begin{itemize}
	\item \texttt{calHn.\textit{l}} ($\texttt{\textit{l}}=0,1,2$) --- Born, one-loop, and two-loop hard massless invariant amplitudes $\widehat{\mathcal{H}}_{n}^{(l)}$, defined in Eq. \eqref{eq:calHn.def}.
	\item \texttt{calHn.smallqq.\textit{l}} ($\texttt{\textit{l}}=0,1,2$) --- small-$q^2$ asymptotics of the corresponding amplitudes.
	\item \texttt{calHn.smallT.\textit{l}} ($\texttt{\textit{l}}=0,1,2$) --- small-$T$ asymptotics of the corresponding amplitudes.
	\end{itemize}
	\item \texttt{Hn/*} --- same as \texttt{calHn/*} but for hard massive invariant amplitudes $\widehat{H}_{n}^{(l)}$.
	\item \texttt{calHmn/}
	\begin{itemize}
	\item \texttt{calHmn.\textit{l}} ($\texttt{\textit{l}}=0,1,2$) --- coefficients $\widehat{\mathcal{H}}_{mn}^{(l)}$ in hard massless leptonic tensor, defined in Eq. \eqref{eq:calHmn.def}.
	\item \texttt{calHmn.smallqq.\textit{l}} ($\texttt{\textit{l}}=0,1,2$) --- small-$q^2$ asymptotics of the corresponding coefficients.
	\item \texttt{calHmn.smallT.\textit{l}} ($\texttt{\textit{l}}=0,1,2$) --- small-$T$ asymptotics of the corresponding  coefficients.
	\end{itemize}
	\item \texttt{Hmn/*} --- same as \texttt{calHmn/*} but for hard massive leptonic tensor $\widehat{H}_{mn}^{(l)}$.
	\item \texttt{incAmn/*} --- same as \texttt{calHmn/*} but for inclusive leptonic tensor $\widehat{A}_{mn}^{inc(l)}$.
\end{itemize}

\section{Conclusion}
The main result of the present paper is the cross section of electron-positron annihilation  into two photons, one of which is virtual, taking into account the second-order radiative corrections, including the emission of one and two soft photons. This cross section can be used in many applications, the most important of which is  the extraction from experimental data on electron-positron annihilation into hadrons of the  value  $R(s)$, which is used for calculation of hadron contribution to the muon anomalous magnetic moment.  The  results obtained can be used  when processing data in experiments of both types: by the radiation return method and by the energy scanning method. In both cases it is necessary to take into account the restrictions imposed by experimental data selections criteria. Since it is impossible to give an universal receipt for taking them into account because of variety of these criteria, we don't consider this question here.

In our results for the process \eeggs the contributions from muon loops, as well as from hadron loops and loops of heavier elementary particles, are not taken into account. These contributions appear in the two-loop approximation and are small in the energy region of main interest for us (when $s\lesssim 1$GeV), since they do not contain logarithmic amplification due to infrared and collinear singularities. In this region these contributions have the same (or smaller, due to the numerical smallness of the loop coefficients) order of magnitude as the uncontrolled, model-dependent contributions.

It should be noted also that the formulas that relate the massive and massless amplitudes rely on the assumption that only hard, collinear and soft region are relevant in the effective theory computation \cite{Becher:2007cu}. Meanwhile, it is known that other regions, e.g., ultra-collinear region, do contribute in separate diagrams, vanishing only in the sum. Therefore, the validity of the massive-massless factorization may, in principle, be questioned and a direct calculation of the amplitudes and cross sections obtained in the present paper would be of undoubted value.

\textit{Note added}:
When this work was prepared for publication, a paper \cite{Badger:2023xtl} was published, where the helicity amplitudes for the process $\ell\bar{\ell}\to \gamma\gamma^*$ were evaluated. A comparison of our results for massless amplitudes with those of Ref. \cite{Badger:2023xtl} would be important, but we leave it for future work as the two results are represented in a very different form.

\section*{Acknowledgments}
The work is supported by the Russian Science Foundation, grant number 22-22-00923.

\appendix
\section{MS bar  and on-shell renormalization schemes}

Unfortunately, there are different definitions of relations between bare and renormalized coupling constants  in the literature, despite the canonical papers \cite{tHooft:1973mfk} and \cite{Bardeen1978}. Therefore we provide here the conventions used in this paper.

First, in  $D =4-2\e$ space-time dimension we keep the bare coupling   $\alpha_0 =\tfrac{e_0^2}{4\pi}$ dimensional, with dimension $(\textit{mass})^\e$, rather than introducing an artificial parameter $\mu_0$ in order to make $\alpha_0$ dimensionless.\footnote{$\mu_0$ was introduced, e.g., in Refs. \cite{Catani:1998bh} and  \cite{Gehrmann2005}.}
As before, we prefer to express the following formulae via
\begin{equation}
	a_0=\frac{\alpha_0}{4\pi}\,, \quad a=\frac{\alpha}{4\pi}\,, \quad \aMS=\frac{\alMS}{4\pi}\,,
\end{equation}
rather than via the corresponding coupling constants.

Then we define the \MSb renormalized coupling via the relation
\begin{equation}
	\aMS = a_0 \left(\frac{ e^{\gamma}\mu^{2}}{4\pi}\right)^{-\e}Z_3^{\MSb}(\aMS )\,,\quad
	Z_3^{\MSb}(\aMS ) =1+\tfrac{\beta_0}{\e}\aMS +\tfrac{\beta_1}{2\e} \aMS ^2 +\mathcal{O}(\aMS^3)\,,
	\label{eq:amsviaa0}
\end{equation}
where $\beta_0=-\frac{4n_f}{3}$, $\beta_1=-4n_f$, and $\gamma=0.577\ldots$ is the Euler constant.

The on-shell renormalization constant $Z_3^{\text{OS}}$, which is the ratio of on-shell and bare coupling constants reads \cite{broadhurst1991gauge}
\begin{equation}
	Z_3^{\text{OS}}=\frac{a}{a_0}=1-\tfrac{4n_f}{3\e}\Gamma(1+\e)\left(\tfrac{4\pi}{m^2}\right)^{\e}a
	-\tfrac{4n_f\left(1+7 \e-4 \e ^3\right)}{\e(2-\e) (1-4 \e^2)}\left[\Gamma(1+\e)\left(\tfrac{4\pi}{m^2}\right)^{\e}a\right]^2 +\mathcal{O}(a^3)
	\label{eq:aviaa0}
\end{equation}
The relations \eqref{eq:amsviaa0} and \eqref{eq:aviaa0} should be treated as exact in $\e$.
Note that in the literature, see for example Ref. \cite{Melnikov2000},  the factor $e^{-\gamma\e} $ in Eq. \eqref{eq:amsviaa0} is sometimes replaced with $\Gamma(1+\e )$. While  in the physical space-time dimension $D = 4$ in the NLO approximation these two conventions lead to the same results, thhis is not so in the NNLO approximation.
Eliminating $a_0$, we obtain the exact in $\e$ relations between $a$ and $\aMS$:
\begin{align}
a=& \left(\frac{e^{\gamma } \mu ^2}{4 \pi }\right)^{\e } \Bigg\{\bar{a}
-\frac{4 n_f}{3 \e }\left(\frac{\mu^{2\e}\Gamma (1+\e ) }{m^{2\e}e^{-\gamma  \e }}-1\right) \bar{a}^2
+\bigg[\frac{16n_f^2}{9\e^2}\left(\frac{\mu^{2\e}\Gamma (1+\e ) }{m^{2\e}e^{-\gamma  \e }}-1\right)^2
\nonumber\\&
-\frac{4n_f\, (1+7 \e-4 \e ^3)}{\e\,(2-\e)(1-4\e^2)}\left(\frac{\mu^{2\e}\Gamma (1+\e ) }{m^{2\e}e^{-\gamma  \e }}\right)^2+\frac{2 n_f }{\e}\bigg] \bar{a}^3+O\left(\bar{a}^4\right)\Bigg\}\,,
\\
\bar{a}=& \left(\frac{e^{\gamma } \mu ^2}{4 \pi }\right)^{-\e } a
+\frac{4  n_f}{3 \e }\left(\frac{\mu^{2\e}\Gamma (1+\e ) }{m^{2\e}e^{-\gamma  \e } }-1\right) \left(\frac{e^{\gamma } \mu ^2}{4 \pi }\right)^{-2 \e }a^2
+\bigg[\frac{16n_f^2}{9\e^2}\left(\frac{\mu^{2\e}\Gamma (1+\e ) }{m^{2\e}e^{-\gamma  \e }}-1\right)^2
\nonumber\\&
+\frac{4n_f\, \left(1+7 \e-4 \e ^3\right) }{\e\,(2-\e)(1-4\e^2)}\left(\frac{\mu^{2\e}\Gamma (1+\e ) }{m^{2\e}e^{-\gamma  \e }}\right)^2
-\frac{2n_f}{\e}\bigg] \left(\frac{e^{\gamma } \mu ^2}{4 \pi }\right)^{-3 \e } a^3+O\left(a^4\right)\,.
\end{align}
At $\e=0$ the above relations simplify to
\begin{align}
	a\stackrel{\e=0}{=}& \bar{a}+\frac{4n_f}{3} \ln \left(\tfrac{m^2}{\mu ^2}\right)  \bar{a}^2+ \left(\frac{16 n_f^2}{9}\ln ^2\left(\tfrac{m^2}{\mu ^2}\right)+4 n_f \ln \left(\tfrac{m^2}{\mu ^2}\right)-15n_f\right) \bar{a}^3+O\left(\bar{a}^4\right)\,,
	\\
	\bar{a}\stackrel{\e=0}{=}&a-\frac{4 n_f}{3}\ln \left(\tfrac{m^2}{\mu ^2}\right) a^2
	+ \left(\frac{16 n_f^2}{9} \ln ^2\left(\tfrac{m^2}{\mu ^2}\right)-4n_f\ln \left(\tfrac{m^2}{\mu ^2}\right)+15n_f\right) a^3+O\left(a^4\right)\,.
\end{align}

Let us also present for completeness the exact in $\e$ expressions for the renormalization constants $Z_{m,l}^{(\text{OS})}$ and $Z_{2,l}^{(\text{OS})}$ ($l=1,2$) defined via
\begin{align}
	m_0&=mZ_m^{(\text{OS})}\,,&\qquad 
	Z_m^{(\text{OS})}&=1+\sum_{l\geqslant1} \left(a_0\tfrac{\Gamma(1+\e)}{(4\pi)^{-\e}m^{2\e}}\right)^l Z_{m,l}^{(\text{OS})}\,,
	\\
	\psi_0&=\sqrt{Z_2^{(\text{OS})}}\psi\,,&\qquad
	Z_2^{(\text{OS})}&=1+\sum_{l\geqslant1} \left(a_0\tfrac{\Gamma(1+\e)}{(4\pi)^{-\e}m^{2\e}}\right)^l Z_{2,l}^{(\text{OS})}\,\,.
\end{align}
We have
\begin{align}
	Z_{m,1}^{(\text{OS})}=& Z_{2,1}^{(\text{OS})}= -\tfrac{3-2 \e}{\e  (1-2 \e)}
	\\
	Z_{m,2}^{(\text{OS})} = &
	\left(\tfrac{\left(2 \e ^3-5 \e ^2-\e +2\right) n_f}{(\e -1) \e ^3 (\e +1)}+\tfrac{(\e +1) (2 \e -1) \left(2 \e ^2-\e -2\right)}{2 (\e -1) \e ^4}\right) 
	\bigg[\, _3F_2\left(1,\tfrac{3}{2}-\e ,\e ;\tfrac{3}{2},3-2 \e ;1\right)
	\nonumber\\
	&+\tfrac{\, _3F_2\left(\tfrac{1}{2},1,2 \e -1;2-\e ,\e +\tfrac{1}{2};1\right)}{1-2\e}\bigg] 
	+\tfrac{3 (\e -1) (2 \e +1) n_f}{\e ^3 (\e +1) (2 \e -1)}+\tfrac{8 \e ^4-9 \e +3}{2 \e ^4 (2 \e -1)^2}
	\nonumber\\
	&-\tfrac{2(\e -1) \left(8 \e ^3-6 \e ^2-5 \e +4\right) \Gamma \left(\tfrac{3}{2}-2 \e \right) \Gamma \left(\e +\tfrac{1}{2}\right) \Gamma (-\e )^2}{4^\e\pi  \e  (2 \e -1)^2 \Gamma (3-3 \e )}\,,
	\\
	Z_{2,2}^{(\text{OS})}=&\tfrac{4-\e}{2}Z_{m,2}^{(\text{OS})}
	-\tfrac{4 \left(4 \e ^2-5\right) n_f}{(\e -1) \e  (2 \e +3) \left(1-4 \e ^2\right)}-\tfrac{(2 \e -3) \left(16 \e ^3+20 \e ^2-4 \e -13\right)}{2 \e  \left(1-4 \e ^2\right)^2}
	\nonumber\\&
	+\tfrac{2 (2 \e -3) \left(8 \e ^3-6 \e ^2-9 \e +3\right) \Gamma \left(\tfrac{3}{2}-2 \e \right) \Gamma \left(\e +\tfrac{3}{2}\right) \Gamma (-\e )^2}{4^\e\pi  \left(1-4 \e ^2\right)^2 \Gamma (2-3 \e )}\,.
\end{align}

The $\e$-expansion of these formulas
\begin{multline}
	Z_{m,2}^{(\text{OS})} = \tfrac{2 n_f+\tfrac{9}{2}}{\e ^2}
	+\tfrac{7 n_f+\tfrac{45}{4}}{\e }
	+\left(\tfrac{69}{2}-16 \zeta _2\right) n_f-30 \zeta _2-12 \zeta _3+48 \zeta _2 \ln{2}+\tfrac{199}{8}
	\\
	+  \Big[ \left(-80 \zeta _2-56 \zeta _3+96 \zeta _2 \ln{2}+\tfrac{463}{4}\right)n_f -192 a_4-165 \zeta _2-132 \zeta _3+252 \zeta _4
	\\
	-96 \zeta _2 \ln^2{2}+288 \zeta _2 \ln{2}+\tfrac{677}{16}-8 \ln^4{2}\Big]\e
	 +\mathcal{O}\left(\e^2\right)
\end{multline}
\begin{multline}
	Z_{2,2}^{(\text{OS})} = 
	\tfrac{4 n_f+\tfrac{9}{2}}{\e ^2}
	+\tfrac{\tfrac{19 n_f}{3}+\tfrac{51}{4}}{\e }
	+\left(\tfrac{1139}{18}-32 \zeta _2\right) n_f-78 \zeta _2-24 \zeta _3+96 \zeta _2 \ln {2}+\tfrac{433}{8}
	\\
	+\Big[-384 a_4+n_f \left(-152 \zeta _2-112 \zeta _3+192 \zeta _2 \ln {2}+\tfrac{20275}{108}\right)-267 \zeta _2-294 \zeta _3+504 \zeta _4
	\\
	-192 \zeta _2 \ln ^2{2}
	+552 \zeta _2 \ln {2}+\tfrac{211}{16}-16 \ln ^4{2}\Big]\e 
	+\mathcal{O}\left(\e ^2\right)
\end{multline}
agree with those presented in Ref.\cite{Melnikov2000}. Here $a_4=\text{Li}_4(1/2)$.

%\bibliographystyle{JHEP}
%\bibliography{eeggstarNNLO}

\providecommand{\href}[2]{#2}\begingroup\raggedright\endgroup

\end{document}